\documentclass[12pt, draftclsnofoot,onecolumn]{IEEEtran}
\usepackage{graphicx,float}
\usepackage{amssymb}
\usepackage{amsmath}
\usepackage{amsfonts}
\usepackage[hang]{subfigure}
\usepackage{epsfig}
\usepackage{xr}
\usepackage{float}
\usepackage{xspace}
\usepackage{bm}
\usepackage{algorithm,algpseudocode}
\usepackage{epstopdf}

\usepackage[font={small}]{caption}
\usepackage{url}
\usepackage{lipsum,multicol}
\usepackage{xcolor}
\usepackage{dsfont}
\usepackage{soul}
\usepackage[compress]{cite}
\usepackage{linegoal}
\usepackage{array}
\usepackage{soul}
\usepackage{multibib}
\newcites{latex}{REFERENCE}

\makeatletter
\newcommand{\algmargin}{\the\ALG@thistlm}
\makeatother
\newlength{\whilewidth}
\algdef{SE}[parWHILE]{parWhile}{EndparWhile}[1]
  {\parbox[t]{\dimexpr\linewidth-\algmargin}{%
     \hangindent\whilewidth\strut\algorithmicwhile\ #1\ \algorithmicdo\strut}}{\algorithmicend\ \algorithmicwhile}%
\algnewcommand{\parState}[1]{\State%
  \parbox[t]{\dimexpr\linewidth-\algmargin}{\strut #1\strut}}

\newlength{\indexwidth}
\algnewcommand{\parNumState}[2]{\State{#1}%
  \parbox[t]{\dimexpr\linewidth-\algmargin-\dimexpr \indexwidth}{\strut #2\strut}}

\algdef{SE}[SUBALG]{Indent}{EndIndent}{}{\algorithmicend\ }%
\algtext*{Indent}
\algtext*{EndIndent}

\ifCLASSOPTIONtwocolumn
\newcommand{\commwidth}{2.6in}
\newcommand{\eqratio}{0.42}
\newcommand{\auswidth}{3.3in}
\newcommand{\auswidthb}{3.5in}
\newcommand{\curvewidth}{3.3in}
\newcommand{\figfourwidth}{3.5in}
\else
\newcommand{\commwidth}{3.2in}
\newcommand{\eqratio}{0.55}
\newcommand{\auswidth}{4in}
\newcommand{\auswidthb}{4.5in}
\newcommand{\curvewidth}{3.5in}
\newcommand{\figfourwidth}{4in}
\fi

\begin{document}
%
\title{Game-Theoretic Multi-Agent Control and Network Cost Allocation under Communication Constraints}
%
%
%
\author{Feier Lian, Aranya Chakrabortty~\IEEEmembership{Senior Member,~IEEE} and Alexandra Duel-Hallen~\IEEEmembership{Fellow,~IEEE}\thanks{The authors are with the Department of Electrical and Computer Engineering, North Carolina State University, Raleigh, NC 27695 USA (e-mail: flian2@ncsu.edu; aranya.chakrabortty@ncsu.edu; sasha@ncsu.edu)} \thanks{Manuscript received May 1, 2016; revised Sept 30, 2016 and Dec 15, 2016. This paper was presented in part at the American Control Conference, Boston Marriott Copley Place, MA, USA, July 6--8, 2016. Financial support from the NSF Grant EECS 1544871 is gratefully acknowledged.}}
\maketitle
\vspace{-10ex}
\begin{abstract}
Multi-agent networked linear dynamic systems have attracted attention of researchers in power systems, intelligent transportation, and industrial automation. The agents might cooperatively optimize a global performance objective, resulting in social optimization, or try to satisfy their own selfish objectives using a noncooperative differential game. However, in these solutions, large volumes of data must be sent from system states to possibly distant control inputs, thus resulting in high cost of the underlying communication network. To enable economically-viable communication, a game-theoretic framework is proposed under the \textit{communication cost}, or \textit{sparsity}, constraint, given by the number of communicating state/control input pairs. As this constraint tightens, the system transitions from dense to sparse communication, providing the trade-off between dynamic system performance and information exchange. Moreover, using the proposed sparsity-constrained distributed social optimization and noncooperative game algorithms, we develop a method to allocate the costs of the communication infrastructure fairly and according to the agents' diverse needs for feedback and cooperation. Numerical results illustrate utilization of the proposed algorithms to enable and ensure economic fairness of wide-area control among power companies. 
\end{abstract}


%
\IEEEpeerreviewmaketitle

\section{Introduction}
%
%
%
%


Multi-agent networked dynamic systems arise in many practical scenarios where the communicating entities are spatially separated or have different economic priorities, e.g., in cyber-physical power networks, multi-vehicle formation, intelligent transportation, industrial automation, etc \cite{sztipanovits2012toward}. Social optimization can be performed when all agents aim to jointly optimize a system-wide objective while a noncooperative differential game is suitable when their objectives are different \cite{basar85}. The linear-quadratic regulator (LQR) optimization objectives are frequently employed in the literature due to their tractability, feasibility of distributed implementation, and broad applicability of the quadratic utility function \cite{basar85,lewis1995optimal,LUKES197196,Mukaidani2006}.

\begin{figure}[!b]
  	\centering
  	\includegraphics[width=\commwidth]{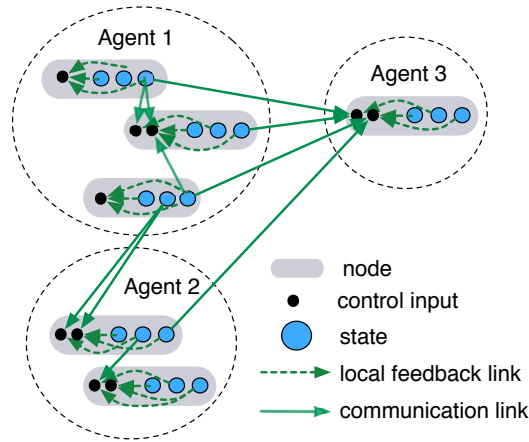}
  	\caption{The communication structure of the multi-agent system.}
  	\label{comm:fig}
  \end{figure}

A networked multi-agent dynamic system with multiple nodes is illustrated in Fig.\ref{comm:fig}. Every agent owns a subset of nodes, where each node contains several states and control inputs. To achieve a desired performance objective, it is often necessary to employ state or output feedback. In this paper, frequently used assumptions of LQR optimization and static state feedback are employed \cite{lewis1995optimal}. Without loss of generality, we define the feedback links from states to control inputs within one node as {\it local feedback links}, which incur negligible expense, and the feedback links across different nodes as {\it communication links}. The traditional state-feedback centralized LQR optimization \cite{lewis1995optimal} and the linear-quadratic games \cite{basar85,LUKES197196,Mukaidani2006} require a dense feedback matrix and, thus, communication links from every state to every control input, which necessitates significant information exchange among the system nodes, and, thus, large communication infrastructure investment to assure desired rate and delay constraints. Due to wide applicability of proposed methods, we do not assume specific communication medium or network topology. Instead, we address the following question: how to reduce the {\it communication cost}, given by the number of communication links (see Fig.\ref{comm:fig}), while maintaining desired control objectives? To answer this question, we develop a family of {\it sparse} designs, which provide a trade-off between the communication cost and control performance, and reveal the most critical communication links. By limiting the number of communicating state/control input pairs, we reduce the overall bill for leasing bandwidth in an existing network or investment in a dedicated communication infrastructure. However, computation of the actual economic benefit depends on specific communication technology and application and is beyond the scope of this paper. 

Sparsity-constrained optimization has been investigated in \cite{bahmani2013greedy,beck2013sparsity}, and state-feedback optimal-control LQR algorithms for sparsity promotion were addressed in \cite{dorjovchebulTPS14,lin2013design}. However, these methods employ global optimization objectives and centralized implementation, which limit their applicability to multi-agent systems. Moreover, distributed approaches in the literature \cite{lamperski2015optimal,mota2013d} cannot accommodate specified sparsity constraints and different optimization objectives of the agents.

In this paper, we investigate LQR optimization for dynamic systems with linear state feedback {\it under the constraint on the communication cost}, i.e. the {\it sparsity} constraint, given by the number of communication links, expressed in terms of the off-diagonal cardinality of the state feedback matrix. First, to solve the {\it centralized} sparsity-constrained optimization problem, we employ the greedy Gradient Support Pursuit (GraSP) algorithm \cite{bahmani2013greedy}, which was shown to provide accurate approximations to sparsity-constrained solutions for a wide class of optimization functions. The proposed method also utilizes the restricted Newton step \cite{fardad2009optimal} to speed up convergence. Second, we develop a {\it noncooperative linear-quadratic game} among the agents, under the global communication cost constraint. To compute a Nash equilibrium of this game, we combine the ideas of GraSP and iterative gradient descent approaches \cite{ratliff2013characterization,li2013designing}. In the resulting algorithm, the computation of the players' (agents') utilities is {\it distributed} and requires limited information exchange. Third, we convert the proposed noncooperative game into a potential game \cite{li2013designing} where the players' utilities agree, thus producing a sparsity-constrained {\it distributed social optimization}. The games developed in this paper can be viewed as {\it Network Formation Games} (NFGs) \cite{van2005models} since players take strategic moves to form a network from states to control inputs. Moreover, using the above algorithms, we apply cooperative NFG theory \cite{van2005models} and the Nash Bargaining Solution (NBS) \cite{Avrachenkov2015265,kawamori2016nash} to allocate the costs of communication among the agents proportionally to the benefits they derive from sparsity-constrained feedback and cooperation. This {\it network cost allocation method} improves on our previous WAC cost allocation approaches \cite{Lian:2014aa,lianensuring}, which employed heuristics, relied on the centralized optimization in \cite{dorjovchebulTPS14}, and extrapolated the costs of a dense network \cite{Mukaidani2006} to sparse scenarios. Finally, we present numerical results for an example of wide-area control (WAC) of power systems, which helps in suppression of inter-area power oscillations, but potentially requires a substantial investment in the communication network needed to exchange state feedback information \cite{naspinet,pramod,lianensuring,dorjovchebulTPS14,deng2012communication,chenine2009survey,pthorp}. These results are shown vs. the sparsity constraint, from dense feedback \cite{Mukaidani2006} to the decentralized implementation \cite{lianensuring}, thus illustrating the trade-off between the communication cost and the control performance.

{\it The main contributions of this paper are:}
\begin{itemize}
\item Development and analysis of centralized and distributed social optimization algorithms and a noncooperative linear-quadratic game for multi-agent LQR optimal control with static linear state feedback under the constraint on the number of communicating state-control input pairs;
\item Development of fair network cost allocation algorithm under sparsity constraints;
\item Enabling sparsity-constrained designs and network cost allocation for a multi-area power system that employs wide-area control.
\end{itemize}

This paper is organized as follows. The system model and the communication-cost-constrained social optimization is presented in Section II. In Section III, the multi-agent system model is developed, and the sparsity-constrained distributed differential games are discussed. Section IV describes the proposed network cost allocation algorithm. In Section V, we present an example of utilizing the proposed methods for WAC of power systems. Numerical results and discussion for the Australian power system example are contained in Section VI. Finally, some concluding remarks are made in Section VII.


\section{System Model and the Communication-Cost-Constrained Centralized Optimization}
\label{cent_sparse:sec}

The linear dynamic system with $n$ nodes illustrated in Fig.\ref{comm:fig} is described by the following state-space equation. 
\begin{equation}
	\dot{\boldsymbol x}(t)={\boldsymbol A}{\boldsymbol x}(t)+{\boldsymbol B}{\boldsymbol u}(t) + {\boldsymbol D}{w}(t),~ \boldsymbol x_0(t) = \boldsymbol 0.
	\label{short:eq}
\end{equation}
\noindent where ${\boldsymbol x}(t){=}({\mathcal X}_1^{\mathrm{T}}(t),...,{\mathcal X}_n^{\mathrm{T}}(t))^{\mathrm{T}}{\in} \mathbb{R}^{m\times 1}$ is the vector of states, $\mathcal{X}_j {\in} \mathbb{R}^{m_j\times 1}$ is the vector of states for node $j{\in}\{1,...,n\}$, $m_j$ is the number of states in node $j$, $m=\sum_{j=1}^n{m_j}$, ${\boldsymbol u}(t)=(\mathcal U_1(t)^{\mathrm{T}},...,\mathcal U_n(t)^{\mathrm{T}})^{\mathrm{T}}\in \mathbb{R}^{q\times 1}$ is the vector of control inputs, $\mathcal{U}_j \in \mathbb{R}^{p_j\times 1}$ is the the vector of control input of node $j$, $p_j$ is the number of control inputs in node $j$, $q=\sum_{j=1}^{n}{p_j}$, $w(t)$ is a scalar impulse disturbance input, and $\boldsymbol A,\boldsymbol B,\boldsymbol D$ are matrices with appropriate dimensions, among which matrix $\boldsymbol{A}$ determines the physical topology of the system \cite{lewis1995optimal}.


We assume linear static feedback is employed, and thus the control input satisfies
\begin{equation}
{\boldsymbol u}(t) = -{\boldsymbol K}\boldsymbol{x}(t)
\label{feedback:eq}
\end{equation}
\noindent where ${\boldsymbol K}\in {\mathbb R}^{q\times m}$ is the feedback gain matrix, with $\boldsymbol{u}(t) = (u_1(t),...,u_q(t))^{\mathrm{T}}$, and $\boldsymbol{x}(t) = (x_1(t),...,x_m(t))^{\mathrm{T}}$. If the coefficient $K_{ij}\neq 0$, the system (shown in Fig.\ref{comm:fig}) contains a communication link that delivers the data of state $j$ to control input $i$. We will refer to the tuple $(x_j(t), u_i(t))$ where $j=1,...,m$, and $i=1,...,q$, as a
{\it state-control input link} in the rest of the paper. Since the states ${\boldsymbol x}(t)$ and the control inputs ${\boldsymbol u}(t)$ are organized according to their physical locations, the matrix $\boldsymbol K$ is in the form
\begin{equation}
\boldsymbol K=\begin{bmatrix}{\boldsymbol K}_{11} & {\boldsymbol K}_{12} & \cdots& {\boldsymbol K}_{1n}\\
					{\boldsymbol K}_{21} & {\boldsymbol K}_{22} & \cdots & {\boldsymbol K}_{2n}\\
					~ & ~ & \vdots & ~\\
					{\boldsymbol K}_{n1} & {\boldsymbol K}_{n2} & \cdots & {\boldsymbol K}_{nn}\\
	\end{bmatrix}
	\label{Kblock:eq}
\end{equation}
where the block ${\boldsymbol K}_{ij}\in \mathbb{R}^{p_i\times m_j}$ represents feedback of the states of node $j$ to the control inputs of node $i$, with $i=j$ corresponding to local feedback and $i\neq j$ --- to communication links (see Fig.\ref{comm:fig}). Without loss of generality, we define the communication cost as the number of communication links associated with the off-diagonal blocks of ${\boldsymbol K}$, given by
\begin{equation}
\mathrm{card}_{\mathrm{off}}({\boldsymbol K})=\sum_{i,j=1,i\neq j}^{n}{\mathrm{nnz}({\boldsymbol K}_{ij})}
\label{cardoff:eq}
\end{equation}
\noindent where $\mathrm{nnz}(\cdot)$ operator counts the number of nonzero elements in a matrix. The proposed algorithms can be easily adapted to other sparsity criteria. Additional notation used in the algorithms is described in Table I.

\begin{table}[!b]
\centering
\caption{Notation used in the Algorithms 1 and 2.}
\begin{tabular}{p{0.15\columnwidth}p{0.75\columnwidth}}
\hline
{\bf Term} & {\bf Definition}\\
\hline
$||{\boldsymbol K}||_2$ & Frobenius norm of the matrix ${\boldsymbol K}$, defined by $\mathrm{trace}({\boldsymbol K}^{\mathrm{T}}{\boldsymbol K})$. \\
\hline
$\mathrm{supp}({\boldsymbol K})$ & The support set of the matrix ${\boldsymbol K}$, i.e., the set of indices of the nonzero entries of matrix ${\boldsymbol K}$ \cite{bahmani2013greedy}.\\
\hline
$[{\boldsymbol K}]_s$ & The matrix obtained by preserving only the $s$ largest-magnitude entries of the matrix ${\boldsymbol K}$, and setting all other entries to zero. \\
\hline 
${\boldsymbol K}^{\mathrm{off}}$ & The matrix obtained by preserving only the off-diagonal blocks of the matrix ${\boldsymbol K}$ (see (\ref{Kblock:eq})) and setting all other entries to zero. \\
\hline
${\boldsymbol K}^{\mathrm{diag}}$ & The matrix obtained by preserving the diagonal blocks of the matrix ${\boldsymbol K}$ and setting all other entries to zero.\\
\hline
$\nabla_{\boldsymbol K}J({\boldsymbol K})$ & The gradient of the scalar function $J({\boldsymbol K})$ with respect to the matrix ${\boldsymbol K}$ \cite{rautert1997computational}. Assuming ${\boldsymbol K}\in \mathbb{R}^{m\times n}$, $\nabla_{\boldsymbol K}J({\boldsymbol K})$ is given by a $m\times n$ matrix with the elements $[\nabla_{\boldsymbol K}J({\boldsymbol K})]_{ij} = \partial J/ \partial K_{ij}$ .\\ 
\hline
$\nabla_{\boldsymbol K}J({\boldsymbol K})|_{\mathcal{T}}$ & The gradient of the scalar function $J({\boldsymbol K})$ with respect to the matrix ${\boldsymbol K}$ projected onto the index set $\mathcal T$. The matrix $\nabla_{\boldsymbol K}J({\boldsymbol K})|_{\mathcal{T}}$ is obtained by preserving only the entries of $\nabla_{\boldsymbol K}J({\boldsymbol K})$ with indices in the set $\mathcal T$ and setting other entries to zero. \\
\hline
$\Delta_{\mathrm{nwt}}({\boldsymbol K},\mathcal T)$ & The restricted Newton step of function $f({\boldsymbol K})$ at matrix ${\boldsymbol K}\in \mathbb{R}^{m\times n}$ under the structural constraint $\mathrm{supp}({\boldsymbol K}) \subset \mathcal{T}$. First, the $mn{\times} 1$ vector $\boldsymbol x$ is computed by stacking the columns of $\boldsymbol K$, and the function $g(\boldsymbol x)$ is defined as $g(\boldsymbol x)\triangleq f(\boldsymbol K)$. Then the $mn{\times} 1$ restricted Newton step vector $\Delta_{\mathrm{nwt}}({\boldsymbol x},\mathcal T)$ of $g(\boldsymbol x)$ at $\boldsymbol x$ \cite{bahmani2013greedy} is computed using the conjugate gradient (CG) method \cite{lin2013design}. The vector $\Delta_{\mathrm{nwt}}({\boldsymbol x},\mathcal T)$ is then converted into an $m{\times} n$ matrix by stacking the consecutive $m {\times} 1$ segments of $\Delta_{\mathrm{nwt}}({\boldsymbol x},\mathcal T)$.\\
\hline
\end{tabular}
\label{notation:tb}
\end{table}

For the model (\ref{short:eq}, \ref{feedback:eq}), the {\it social global} LQR objective function is given by \cite{lewis1995optimal}
\begin{equation}
J(\boldsymbol{K})=\int_{t=0}^{\infty}{[{\boldsymbol x}(t)^T {\boldsymbol Q}{\boldsymbol x}(t)+ {\boldsymbol u}(t)^T{\boldsymbol R}{\boldsymbol u}(t) ]dt}
\label{J:eq}
\end{equation}
\noindent where ${\boldsymbol Q}$ and ${\boldsymbol R}$ are the positive semidefinite and positive definite design matrices with dimensions $m\times m$ and $q\times q$, respectively. The minimization of (\ref{J:eq}) with respect to $\boldsymbol{K}$ results in dense feedback, i.e., communication links from every state to every control input.

Next, we formulate the {\it social optimization under the communication cost constraint $s$}: 
\begin{eqnarray}
&\underset{\boldsymbol K}{\mathrm{min}}& J(\boldsymbol{K})\nonumber\\
&\mbox{s.t.}&  \mathrm{card}_{\mathrm{off}}({\boldsymbol K}) \leq s \nonumber\\
&&\dot{\boldsymbol x}(t)={\boldsymbol A}{\boldsymbol x}(t)+{\boldsymbol B}{\boldsymbol u}(t) + {\boldsymbol D}{w}(t)\nonumber\\
& & {\boldsymbol u}(t) = -{\boldsymbol Kx}(t)
\label{sparse_cent:eq}
\end{eqnarray}
\noindent
The optimization (\ref{sparse_cent:eq}) produces a system with at most $s$ state-control input links. Direct solution of (\ref{sparse_cent:eq}) can have combinatorial complexity \cite{bahmani2013greedy}, so we utilize a numerically efficient {\it GraSP} method \cite{bahmani2013greedy} in the proposed {\it centralized Algorithm \ref{cent:alg}}. Given the overall sparsity constraint $s$, in each iteration of Step 2(1--4), the algorithm extends the matrix ${\boldsymbol K}$ along its steepest $2s$ gradient-descent directions. In Step 2(1), the gradient of $\boldsymbol J(\boldsymbol K)$ over $\boldsymbol K$ is computed as \cite{fardad2009optimal} 
\begin{equation}
\nabla_{{\boldsymbol K}}J({\boldsymbol K}) = 2(\boldsymbol R \boldsymbol K - \boldsymbol B^{\mathrm{T}}\boldsymbol P)\boldsymbol L
\label{gradJ:eq}
\end{equation}
\noindent where the matrices $\boldsymbol P$ and $\boldsymbol L$ are the unique solutions of the following Lyaponuv equations
\begin{align}
\label{lyapJ:eq}
&(\boldsymbol A - \boldsymbol B\boldsymbol K)^{\mathrm{T}}\boldsymbol P + \boldsymbol P(\boldsymbol A - \boldsymbol B\boldsymbol K) + \boldsymbol Q + \boldsymbol K^{\mathrm{T}}\boldsymbol R \boldsymbol K^{\mathrm{T}}=0\nonumber\\
& (\boldsymbol A - \boldsymbol B\boldsymbol K)\boldsymbol L + \boldsymbol L(\boldsymbol A - \boldsymbol B\boldsymbol K)^{\mathrm{T}} + \boldsymbol D \boldsymbol D^{\mathrm{T}} = 0
\end{align}
\noindent In Step 2(4), the matrix $\boldsymbol K$ is updated using the restricted Newton step, and the step size $\lambda$ is chosen via the Armijo line search \cite{Boyd:2004aa}. In Step 2(5), pruning is performed to impose the constraint $s$. To guarantee stability of the feedback matrix after pruning, we provide a backtracking option to return to a previously found stable solution, which has $s-1$ or fewer communication links. (The stopping criterion in Step 2(6) was also used to determine convergence in the sparsity-promotion algorithm \cite{dorjovchebulTPS14}.) Note that in Step 2(6), $q$ is the total number of control inputs, and $m$ is the total number of states. Note also that as $s$ grows, the algorithm retains the links that are the most {\it critical} for minimizing the objective (\ref{J:eq}).


\begin{algorithm}[!t]
	 \caption{Minimizing the centralized LQR objective under the global communication cost constraint $s$.}
	 \label{cent:alg}
	\begin{algorithmic}
	\State 1 {\it Initialization} 
	\State ${\boldsymbol K} := {\boldsymbol K}_{0}$
	\State 2 {\it Iteration} 
	\While {stopping criteria are not met}
	\parNumState {(0) }{${\boldsymbol K}^{\mathrm{prev}} := {\boldsymbol K}$}
	\parNumState {(1) } {Compute gradient of $J(\boldsymbol K)$ w.r.t ${\boldsymbol K}$: ${\boldsymbol g} = \nabla_{{\boldsymbol K}}J({\boldsymbol K})$}
	\parNumState {(2) }{Identify up to $2s$ off-diagonal block directions: $\mathcal{Z} = \mathrm{supp}([{\boldsymbol g}^{\mathrm{off}}]_{2s})$}
	\parNumState {(3) } {Merge support: $\mathcal{T} = \mathcal{Z} \cup \mathrm{supp}({\boldsymbol K}).$}
	\parNumState {(4) }{Descend using the Newton step of $J$ restricted to $\mathcal{T}$: ${\boldsymbol K} := {\boldsymbol K} + \lambda \Delta_{\mathrm{nwt}}({\boldsymbol K},\mathcal{T}).$}
	\parNumState {(5) } {Prune communication links: ${\boldsymbol K}:={\boldsymbol K}^{\mathrm{diag}}+[{\boldsymbol K}^{\mathrm{off}}]_{s}$}
    \parNumState {(6) } {Stopping criterion: \\$||{\boldsymbol K}-{\boldsymbol K}^{\mathrm{prev}}||_2 < \epsilon_{\mathrm{abs}}\sqrt{qm}+\epsilon_{\mathrm{rel}}||{\boldsymbol K}^{\mathrm{prev}}||_2$}
    \EndWhile
	\State 3 {\it Polishing}
	\State {$\mathcal I = \mathrm{supp}({\boldsymbol K})$} 
	\While{not $||\nabla_{\boldsymbol K} J|_{{\mathcal I}}||_2/\sqrt{qm}< \epsilon_2$ }
	\State Descend using the Newton step of $J$ restricted to ${\mathcal I}$: \\${\boldsymbol K} := {\boldsymbol K} + \lambda \Delta_{\mathrm{nwt}}({\boldsymbol K},\mathcal{I}).$
	\EndWhile
	 \end{algorithmic}
\end{algorithm}

\section{Multi-Agent System Model and Communication-Cost-Constrained Linear-Quadratic Games}
\label{densegame:sec}
Suppose that there are $r$ agents, where the agent $i$ owns $n_i$ nodes in the system (\ref{short:eq}) as shown in Fig.\ref{comm:fig}. Without loss of generality, the nodes are partitioned as follows

\begin{gather}
\resizebox{\eqratio\textwidth}{!}{
$\begin{aligned}
\mathcal S_1 = \{1,2,...,n_1\} & \Rightarrow &\mbox{belongs to agent $1$.}\\
\mathcal S_2 = \{n_1+1,n_1+2,...,n_1+n_2\} & \Rightarrow &\mbox{belongs to agent $2$.}\\
\hdots~ & \Rightarrow & \hdots~ \\
\mathcal S_r = \{n_1+n_2+...+n_{r-1}+1,...,n\} & \Rightarrow &\mbox{belongs to agent $r$.}
\label{partition:eq}
\end{aligned}$}
\end{gather} 

We can rewrite the states and control inputs of each agent $i$ in (\ref{short:eq}) as follows:
\begin{equation}
\dot{ \boldsymbol x}_i(t)= \sum_{k=1}^{r}{ \boldsymbol A_{ik} \boldsymbol x_k(t) } + \sum_{k=1}^{r}{ \boldsymbol B_{ik} \boldsymbol u_k(t)} + \boldsymbol D_i w(t)
\label{big:eq}
\end{equation}
\begin{equation}
\begin{bmatrix}{\boldsymbol u}_1(t)\\ \vdots \\{\boldsymbol u}_r(t) \end{bmatrix}
 = -\begin{bmatrix} {\boldsymbol K}^1 \\ \vdots \\ {\boldsymbol K}^r \end{bmatrix}\cdot {\boldsymbol x}(t)
 \label{K^i:eq}
\end{equation}
\noindent where ${\boldsymbol x}_i(t) = ({\mathcal X}^{\mathrm{T}}_{n_1+...+n_{i-1}+1}(t),..., {\mathcal X}^{\mathrm{T}}_{n_1+...+n_i}(t))^{\mathrm{T}}$$ \in \mathbb{R}^{M_i\times 1}$ is the vector of states for agent $i$, with $ M_i \allowbreak =\allowbreak \sum_{j=n_1 +... +n_{i-1}+1}^{n_1+...+n_i}{m_j}$; ${\boldsymbol u}_i(t)\allowbreak {=}\allowbreak[{\mathcal U}^{\mathrm{T}}_{n_1{+}...{+}n_{i-1}{+}1}(t),\allowbreak \hdots, {\mathcal U}^{\mathrm{T}}_{n_1{+}...{+}n_i}(t)]^{\mathrm{T}} \in \mathbb{R}^{N_i\times 1}$ is the vector of control inputs for agent $i$, with $N_i {=} \sum_{j=n_1+...+n_{i-1}+1}^{n_1+...+n_i}{p_j}$, ${\boldsymbol K}^i$ is the submatrix of ${\boldsymbol K}$ associated with the control inputs of the agent $i$, $\boldsymbol D_i$ is the control matrix for the disturbance input that enters agent $i$, and $\boldsymbol D=\mathrm{col}(\boldsymbol D_1,...,\boldsymbol D_r)$.

Next, we briefly summarize results on linear-quadratic games for this system \cite{basar85,Mukaidani2006}. The agents in (\ref{partition:eq}) are viewed as players that optimize their individual, or selfish, objectives $J_i$ by selecting their control inputs ${\boldsymbol u}_i(t)$, for $i=1,...,r$. The {\it selfish} objective of the player $i$ is given by
\begin{equation}
J_i(\boldsymbol{u}_1, {\boldsymbol u}_2,..., {\boldsymbol u}_r){=}\int_{t=0}^{\infty}{[{\boldsymbol x}(t){\boldsymbol Q}_i{\boldsymbol x}(t)^{\mathrm{T}} {+} {\boldsymbol u}_i(t)^{\mathrm{T}}{\boldsymbol R}_i {\boldsymbol u}_i(t)]dt}
\label{Ji:eq} 
\end{equation}

\noindent where ${\boldsymbol R}_i {\in} \mathbb{R}^{N_i \times N_i}$, and ${\boldsymbol Q}_i{\in} \mathbb{R}^{m\times m}$ are positive semidefinite and positive definite matrices, respectively, chosen to improve the $i^{\mathrm{th}}$ user's objective. A {\it Nash Equilibrium (NE)} is achieved when it is impossible for any player to improve its objective function by unilaterally changing its strategy. At a given NE, the players employ Nash strategies $({\boldsymbol u}_1^*(t),{\boldsymbol u}_2^*(t),...,{\boldsymbol u}_r^*(t))$ defined as \cite{basar85}
\begin{equation}
J_i(\boldsymbol{u}_i^*(t), \boldsymbol{u}_{-i}^*(t)) {\leq} J_i(\boldsymbol{u}_i(t), \boldsymbol{u}_{-i}^*(t)) ,\forall \boldsymbol u_i(t),t {=} [0,\infty)
\label{NE:eq}
\end{equation}
\noindent for $\forall i{\in} \{1,...r\}$, where $\boldsymbol{u}_{-i}(t){:=} (\boldsymbol{u}_1(t), \hdots, \boldsymbol{u}_{i{-}1}(t),\boldsymbol{u}_{i+1}(t),\allowbreak \hdots, \boldsymbol{u}_{r}(t))$ is the tuple of strategies formed by all players except for the player $i$. When state feedback is employed (\ref{K^i:eq}), the Nash strategies $\boldsymbol{u}_i^*(t)$ in eq.(\ref{NE:eq}) can be determined by solving the cross-coupled algebraic Riccati equations (CARE) (eq.(8) in \cite{Mukaidani2006}), where the solution to CARE exists and is unique when the system is weakly coupled. However, CARE produces a dense feedback matrix \cite{Mukaidani2006}, and, thus, communication links from every state to every control input.

To limit the communication cost in a noncooperative scenario, we formulate a {\it linear-quadratic noncooperative NFG} where the players can establish at most $s$ state-control input communication links (see Fig.\ref{comm:fig}) while each player aims to minimize its selfish LQR objective (\ref{Ji:eq}). A Nash Equilibrium of this game results in a communication network with cost bounded by $s$. Thus, Nash strategies $({\boldsymbol u}_1^*(t), {\boldsymbol u}_2^*(t),..., {\boldsymbol u}_r^*(t))$ satisfy for $\forall i\in\{1,...,r\}$
\begin{align}
&J_i(\boldsymbol{u}_i^*(t), \boldsymbol{u}_{-i}^*(t)) \leq J_i(\boldsymbol{u}_i(t), \boldsymbol{u}_{-i}^*(t))~~,\forall {\boldsymbol u}_i(t) \nonumber\\
\mbox{s.t. } & \mathrm{card}_{\mathrm{off}}({\boldsymbol K}) \leq s 
\label{sparse_NE:eq}
\end{align}
\noindent Equivalently, since linear static feedback (\ref{K^i:eq}) is employed, the strategy of player $i$ is given by the submatrix ${\boldsymbol K}^i$ in (\ref{K^i:eq}), and Nash strategies are expressed as $({{\boldsymbol K}^1}^*,{{\boldsymbol K}^2}^*,...,{{\boldsymbol K}^r}^*)$, which satisfy for each $i\in\{1,...,r\}$
\begin{align}
&J_i({{\boldsymbol K}^{i}}^*,{{\boldsymbol K}^{-i}}^*) \leq J_i({\boldsymbol K}^{i},{{\boldsymbol K}^{-i}}^*) ~,\forall {\boldsymbol K}^{i}\nonumber\\
\mbox{s.t. } & \mathrm{card}_{\mathrm{off}}({\boldsymbol K}) \leq s 
\label{NE_sparseK:eq}
\end{align}
\noindent where the tuple ${\boldsymbol K}^{-i}:= ({\boldsymbol K}^{1},...,{\boldsymbol K}^{i-1}\allowbreak,\allowbreak {\boldsymbol K}^{i+1}\allowbreak,\allowbreak...,{\boldsymbol K}^{r})$ represents the strategies of all other players except $i$. In this case, as $s$ increases, the optimization (\ref{NE_sparseK:eq}) retains the most critical state-control input links needed for the noncooperative optimization, characterized by the set of selfish objectives (\ref{Ji:eq}).

The proposed game is described in {\it Algorithm \ref{graspNE:alg}}. It is inspired by the iterative gradient descent methods in \cite{ratliff2013characterization,li2013designing}, where each player takes a small step towards minimizing its own objective while other players' strategies are fixed. In each step associated with player $i$, we use the GraSP algorithm \cite{bahmani2013greedy} to update the strategic variable ${\boldsymbol K}^i$ while maintaining the overall sparsity constraint. Thus, in the submatrix ${\boldsymbol K}^i$ (\ref{K^i:eq}), the elements representing local feedback, i.e., those in the blocks ${\boldsymbol K}_{jj} \in \mathbb{R}^{p_j\times m_j}$ in (\ref{Kblock:eq}) for $j\in \mathcal S_i$, are free variables while the off-diagonal blocks in (\ref{Kblock:eq}) are subject to the sparsity constraint. The computation of the gradient of $J_i$ w.r.t. player $i$'s strategy $\boldsymbol K^{i}$ is similar to that in (\ref{gradJ:eq}--\ref{lyapJ:eq}).
\begin{align}
\nabla_{{\boldsymbol K}^i}J_i({\boldsymbol K}^i,{\boldsymbol K}^{-i}) = 2(\boldsymbol R_i\boldsymbol K^{i} - \boldsymbol D^{\mathrm{T}}\bar{\boldsymbol P_i})\bar{\boldsymbol L_i}
\end{align}
\noindent where $\bar{\boldsymbol P_i}$ and $\bar{\boldsymbol L_i}$ are the unique solutions of the following Lyapunov equations
\begin{align}
&(\bar{\boldsymbol A}_i - \boldsymbol B_i\boldsymbol K^{i})^{\mathrm{T}}\bar{\boldsymbol P_i} + \bar{\boldsymbol P_i}(\bar{\boldsymbol A}_i - \boldsymbol B_i\boldsymbol K^{i}) + \boldsymbol Q_i + {\boldsymbol K^{i}}^{\mathrm{T}}\boldsymbol R_i \boldsymbol K^i = 0\nonumber\\
&(\bar{\boldsymbol A}_i - \boldsymbol B_i\boldsymbol K^{i})\bar{\boldsymbol L_i} + \bar{\boldsymbol L_i}(\bar{\boldsymbol A}_i - \boldsymbol B_i\boldsymbol K^{i})^{\mathrm{T}} + \boldsymbol D\boldsymbol D^{\mathrm{T}} = 0
\end{align}
\noindent and $\bar{\boldsymbol A_i}=\boldsymbol A -\sum_{j=1,j\neq i}^{r}{\boldsymbol B_j\boldsymbol K^j}$, $\boldsymbol B_j = \mathrm{col}(\boldsymbol B_{1j},...,\boldsymbol B_{rj})$.

\begin{algorithm}[!b]
	 \caption{Noncooperative game under the global communication cost constraint $s$.}
	\begin{algorithmic}
	\State {1 {\it Initialization}} 
	\parState{${\boldsymbol K} := {\boldsymbol K}_0$}
    \parState{Initialize the link constraint $s_i$ for each player $i~,i=1,...,r, ~\sum_{i=1}^r{s_i}=s$.}
	\State 2 {\it Iteration} 
	\While {stopping criteria are not met}
	\State{${\boldsymbol K}^{\mathrm{prev}}:= {\boldsymbol K}$}
	\For {$i$ = $1 \hdots r$}
	\parNumState {(1) } {if $i>1$, $s_i = s-\mathrm{card}_{\mathrm{off}}({\boldsymbol K}^{-i}) $.}
	\parNumState {(2) } {Compute gradient of $J_i$ in (\ref{Ji:eq}) w.r.t ${\boldsymbol K}^i$:\\ ${\boldsymbol g}_i = \nabla_{{\boldsymbol K}^i}J_i({\boldsymbol K}^i,{\boldsymbol K}^{-i})$ }
	\parNumState {(3) } {Identify up to $2s_i$ off-diagonal block directions: $\mathcal{Z} = \mathrm{supp}([{\boldsymbol g}_i^{\mathrm{off}}]_{2s_i})$}
	\parNumState {(4) } {Merge support: $\mathcal{T} = \mathcal{Z} \cup \mathrm{supp}({\boldsymbol K}^i).$}
	\parNumState {(5) } {Descend using the Newton step of $J_i$ restricted to $\mathcal{T}$: ${\boldsymbol K}^i := {\boldsymbol K}^i + \lambda \Delta_{\mathrm{nwt}}({\boldsymbol K}^i,\mathcal{T}).$}
	\parNumState {(6) } {Prune among the off-diagonal block elements: ${\boldsymbol K}^i:= {\boldsymbol K}^{i,\mathrm{diag}}+[{\boldsymbol K}^{i,\mathrm{off}}]_{s_i} $}
	\EndFor
	\parState {Construct global feedback matrix: ${\boldsymbol K} = [{\boldsymbol K}^1; ... ;{\boldsymbol K}^r]$.}
    \parState {Stopping criteria: \\$||{\boldsymbol K}-{\boldsymbol K}^{\mathrm{prev}}||_2 < \epsilon_{\mathrm{abs}}\sqrt{qm}+\epsilon_{\mathrm{rel}}||{\boldsymbol K}^{\mathrm{prev}}||_2$.}
    \EndWhile
	\State 3 {\it Polishing}
	\State $\mathcal{I}_i = \mathrm{supp}({\boldsymbol K}^i)$, for $i=1,...,r$.
	\While{not $\nabla_{{\boldsymbol K}^i} J_i|_{{\mathcal I}_i}< \epsilon_2~\forall i=1,...,r$ }
	\For{$i=1,...,r$}
	\parState {Descend using the Newton step of $J_i$ restricted to ${\mathcal I}_i$: ${\boldsymbol K}^i := {\boldsymbol K}^i + \lambda \Delta_{\mathrm{nwt}}({\boldsymbol K}^i,\mathcal{I}_i).$}
	\EndFor
	\EndWhile
	 \end{algorithmic}
	 \label{graspNE:alg}
\end{algorithm}

Algorithm \ref{graspNE:alg} is {\it distributed} in the sense that player $i$ individually updates its strategic variable ${\boldsymbol K}^i$. Each player has prior knowledge of the underlying physical system (i.e., the matrices ${\boldsymbol A},{\boldsymbol B},\boldsymbol D$ in (\ref{short:eq})) and broadcasts its strategic move (the submatrix ${\boldsymbol K}^i$) after its strategy is updated at the completion of Step 2(6) and each inner loop of Step 3. 

Finally, note that the social optimization under the sparsity constraint (\ref{sparse_cent:eq}) can also be implemented in a distributed fashion using a potential game \cite{li2013designing}, obtained from (\ref{NE_sparseK:eq}) by replacing the individual objectives in (\ref{NE_sparseK:eq}) with the common social objective (\ref{J:eq}) while the players' strategies are still defined as their control vectors. We refer to this game as {\it distributed social optimization}, and its equilibria are defined as
\begin{align}
&J({{\boldsymbol K}^{i}}^*,{{\boldsymbol K}^{-i}}^*) \leq J({\boldsymbol K}^{i},{{\boldsymbol K}^{-i}}^*) ~,\forall {\boldsymbol K}^{i}\nonumber\\
\mbox{s.t. } & \mathrm{card}_{\mathrm{off}}({\boldsymbol K}) \leq s 
\label{NE_J:eq}
\end{align}
\noindent To compute (\ref{NE_J:eq}), we use Algorithm 2, where the social objective $J()$ replaces $J_i()$ in Step 2(2) and in the Polishing Step 3. 

Finally, we discuss the NEs of the proposed games. Since the LQR objective (\ref{J:eq}) is not convex in the feedback matrix $\boldsymbol{K}$ in general \cite{rautert1997computational}, the noncooperative game (\ref{sparse_NE:eq}, \ref{NE_sparseK:eq}) is not guaranteed to admit a pure-strategy NE \cite{pang2011nonconvex}. On the other hand, the distributed social optimization (\ref{NE_J:eq}) is an exact potential game, and, thus, a pure NE exists for this game \cite{li2013designing}. Moreover, the optimal solution of the social optimization problem (\ref{sparse_cent:eq}) constitutes a NE of (\ref{NE_J:eq}) although the converse does not necessarily hold due to nonconvexity of the LQR objective. Finally, while CARE has a unique NE under the weakly-coupled system assumption \cite{Mukaidani2006}, the cost-constrained games (\ref{sparse_NE:eq}), (\ref{NE_J:eq}) can have multiple NEs.

 \section{Network Cost Allocation}
\label{network:sec}

\begin{algorithm}[!b]
	 \caption{Network Cost Allocation under the Communication Cost Constraint $s$.}
	 \label{alg3:alg}
	\begin{algorithmic}
	\State 1 {\it Find social optimized objective $J_{\mathrm{soc}}(s)$} 
	\Indent
	\parState {$J_{\mathrm{soc}}(s)$ is the optimized objective of (\ref{sparse_cent:eq}) or (\ref{NE_J:eq}) under constraint $s$;}
	\EndIndent
	\State 2 {\it Find decoupled optimized objectives at DNE.} 
	\Indent
	\parNumState{(1) } {The NE (\ref{NE_sparseK:eq}) for $s=0$ is the {\it decoupled} NE (DNE);}

	\parNumState{(2) } {{\it Optimized selfish objective} at {\it DNE}:$J_i^{\mathrm{D}},i = 1,..., r.$}
	\parNumState{(3) } {{\it Total objective at DNE:} ${\tilde J}^{\mathrm{D}} := \sum_{i=1}^{r}{J_i^{\mathrm{D}}}$.}

	\EndIndent
	\State 	3 {\it Find coupled optimized objectives at CNE}
	\Indent
	\parNumState{(1) } {The NE (\ref{NE_sparseK:eq}) under constraint $s$ is {\it coupled} NE ($\mathrm{CNE}(s)$);}

	\parNumState{(2) } {{\it Optimized selfish objective at CNE}:
	$J_i^{\mathrm{C}}(s), ~i = 1,...,r$}

	\parNumState{(3) } {{\it Total objective at $\mathrm{CNE}(s)$}: ${\tilde J}^{\mathrm{C}}(s) = \sum_{i=1}^{r}{J_i^{\mathrm{C}}(s)}.$}

	\EndIndent
	\State 4 {\it Compute and allocate payoffs and network costs} 
	\Indent
	\parNumState{(1) } {The {\it social payoff}: $v_{\mathrm{soc}}(s) = {\tilde J}^{\mathrm{D}} - J_{\mathrm{soc}}(s)$}

	\parNumState{(2) } {The {\it selfish payoffs} (the {\it disagreement point}):\\
	 $v_i(s) = J_i^{\mathrm{D}} - J_i^{\mathrm{C}}(s), ~i = 1, ..., r$;}

    \parNumState{(3) } {The {\it allocated payoffs}:\\
     $\alpha_i(s)=v_i(s)+ \frac{v_{\mathrm{soc}}(s) - \sum_{k=1}^r{v_k(s)}}{r}$}

     \parNumState{(4) } {{\it Proportional allocation of the network cost}:\\ $C_i(s) = \alpha_i(s)/\sum_{i=1}^{r}􏰌{\alpha_i(s)}, ~i=1,...,r.$}
   
    \EndIndent
	 \end{algorithmic}
\end{algorithm}

First, we review relevant results for cooperative NFGs with transferable utility \cite{Peters2008} where the utility of a coalition is viewed as monetary value, which is distributed among the players. Suppose $r$ players cooperatively form a network with the objective of maximizing their payoffs. Several approaches to fair payoff allocation have been proposed in the literature \cite{saad2009distributed,myerson1980conference,Avrachenkov2015265,kawamori2016nash}. We employ the Nash Bargaining Solution (NBS) due to its computational efficiency in NFGs \cite{Avrachenkov2015265,kawamori2016nash}. The NBS payoff allocation algorithm proceeds in {\it three steps} based on \cite{kawamori2016nash}: 

(1) The players cooperate to construct a network that maximizes the {\it global social payoff} $v_{\mathrm{soc}}$; 

(2) The {\it disagreement point} is computed as
\begin{equation}
	\label{dis:eq}
{\bold v} = (v_1,v_2 ..., v_r)
\end{equation}
\noindent where the {\it selfish payoff} $v_i$ is the minimum payoff that the $i^{\mathrm{th}}$ player is willing to accept. For example, (\ref{dis:eq}) can be computed as a NE of a noncooperative NFG where each player aims to satisfy its own selfish objective; 

(3) The overall payoff $v_{\mathrm{soc}}$ is split among the players, with the {\it allocated payoff} of player $i$ given by \cite{kawamori2016nash}
	\begin{equation}
		\label{nbs:eq}
		\alpha_i=v_i + \frac{v_{\mathrm{soc}} - \sum_{k=1}^r{v_k}}{r}.
		\end{equation}

\noindent Note that bargaining is successful when the payoff of the social network is at least as large as the sum of the selfish payoffs, i.e.,
\begin{equation}
	\label{eqB:eq}
	v_{\mathrm{soc}} \geq \sum_{i=1}^{r}{v_i}
	\end{equation}
\noindent or equivalently, each player's allocated payoff is at least as large as its selfish payoff:
\begin{equation}
	\label{delta:eq}
    \xi =\alpha_i -  v_i \geq 0.
\end{equation}
\noindent From (\ref{delta:eq}), all players benefit equally by forming the social network.

Next, we describe the proposed {\it network cost allocation method for sparsity-constrained multi-agent dynamic systems}, which is summarized in {\it Algorithm \ref{alg3:alg}}. Given the global communication cost constraint $s$, in steps 1--3, three different scenarios are analyzed, depending on whether the agents employ: {\it (i)} non-local feedback, i.e., {\it communication bounded by cost $s$}; and {\it (ii) cooperation}. In Step 1, both {\it(i)} and {\it(ii)} are assumed, resulting in the social optimization, implemented using Algorithm \ref{cent:alg} or \ref{graspNE:alg} under the constraint $s$. In contrast, in Step 2 neither {\it (i)} nor {\it (ii)} are employed, and thus the agents play the {\it decoupled} noncooperative game, i.e., (\ref{sparse_NE:eq}) with $s=0$, which results in a {\it decentralized} system using the Polishing Step 3 of Algorithm \ref{graspNE:alg} (see \cite{lianensuring}). Finally, in Step 3, only {\it (i)} is employed, resulting in the {\it coupled} game, i.e., (\ref{sparse_NE:eq}) with the global constraint $s$, implemented using Algorithm \ref{graspNE:alg}. Note that Steps 1 and 3 optimize systems with (at most) $s$ communication links (see Fig.\ref{comm:fig}), while Step 2 restricts communication to local feedback links.  

In Steps 4(1) and 4(2), the {\it payoffs} $v_{\mathrm{soc}(s)}$ and $v_i(s)$ represent the improvement in control performance, or {\it objective reduction}, provided by communication bounded by cost $s$ with (in Step 1) and without (in Step 3) cooperation, respectively, relative to the decentralized implementation (Step 2). It is reasonable to model the minimum payoff an agent $i$ expects from communication by its selfish objective reduction, or {\it selfish payoff}, $v_i(s)$, thus forming the {\it disagreement point}. The {\it allocated payoffs} $\alpha_i(s)$ are computed in Step 4(3). Note that the NBS algorithm interprets the {\it payoffs} $\alpha_i(s)$ and $v_i(s)$ as {\it values} (e.g., monetary payoffs) that agent $i$ derives from communication with and without cooperation, respectively (see (\ref{dis:eq}, \ref{nbs:eq})). If cooperation does not degrade the minimum acceptable value $v_i(s)$, i.e., if (\ref{eqB:eq}, \ref{delta:eq}) hold, then bargaining is successful, and the agents agree to use the links formed by the social optimization found in Step 1, which provides them with the social payoff $v_\mathrm{soc}(s)$ and the individual payoffs $\alpha_i(s),~i=1...r$. The latter condition is satisfied for a practical scenario where, in (\ref{J:eq}) and (\ref{Ji:eq}), the matrix ${\boldsymbol R}$ is block-diagonal with diagonal blocks given by $\boldsymbol{R}_i$, $i=1,...,r$, and
\begin{equation}
\sum_{i=1}^{r}{\boldsymbol{Q}_i} = \boldsymbol{Q}
\label{sumQ:eq}
\end{equation}
\noindent
In this case, $\sum_{i=1}^{r}{J_i(\boldsymbol{K})} = J(\boldsymbol{K}) \footnote{$J_i(\boldsymbol{K})$ is equivalent to (\ref{Ji:eq}). Since we use full state feedback, and the control input $\boldsymbol{u}(t) = -\boldsymbol {Kx}(t)$ is equivalent to (\ref{K^i:eq}), the objective $J_i(\boldsymbol{u}_1, {\boldsymbol u}_2,..., {\boldsymbol u}_r)$ is a function of $\boldsymbol{K}$ and can be expressed as $J_i(\boldsymbol{K})$.},~\forall \boldsymbol{K}$, and thus a NE of the noncooperative game (\ref{NE:eq}) is a feasible solution of the social optimization (\ref{sparse_cent:eq}) for the same value of constraint $s$, resulting in $J_{\mathrm{soc}}(s) \leq \tilde J^{\mathrm{C}}(s)$ (see Step 1 and 3(3)), which is equivalent to (\ref{eqB:eq}). 

We assume that the grand coalition forms in the proposed cooperative game among the agents. This assumption justifies Step 1 in Alg. 3, where all players cooperate. A sufficient condition for this assumption is efficiency of the grand coalition \cite{Hafalir2007242}. We show in \cite{Lian:aa} that a practical coalition-level identity similar to (\ref{sumQ:eq}) guarantees efficiency of the grand coalition.

If cooperation is successful, the feedback data required for implementing the {\it social control strategy} (as specified by the optimal feedback matrix in (\ref{sparse_cent:eq}) or (\ref{NE_J:eq})) will be delivered by the communication network specific to the given medium and application, where {\it the constraint $s$ is chosen based on the desired performance/cost tradeoff}. In the final Step 4(4), we compute each agent's share of the cost of this network proportionally to the allocated payoffs $\alpha_i(s)$. Note that if $\xi(s)$ in (\ref{delta:eq}) is small relative to the average of the selfish payoffs $v_i(s)$ (or $J_{\mathrm{soc}}(s) \approx \tilde J^{\mathrm{C}}(s)$), the benefit of cooperation {\it (ii)} is small relative to that of communication {\it (i)}, and the agents' costs $C_i(s)$ can vary significantly due to the agents' diverse needs for feedback and cooperation as is illustrated in Section \ref{num:sec} and in \cite{lianensuring}. On the other hand, when $\xi(s)$ is relatively large, i.e., $\alpha_i(s)\approx \xi(s)$ for all $i$, the payoffs equalize (\ref{delta:eq}), resulting in equally split cost of the communication network among the agents.

\section{Example: Sparsity-Constrained Wide-Area Control of Power Systems}

\label{WAC:sec}
We validate our algorithms using the example of wide-area control (WAC) of large-scale power system networks. In recent literature such as \cite{dorjovchebulTPS14,pramod}, WAC has been shown to be very useful for suppressing low-frequency oscillations following small-signal disturbances in large power grids, at the cost of communicating real-time data from sensors at one operating region to controllers at others. The sensors are referred to as Phasor Measurement Units (PMUs), all of which are synchronized to each other via GPS. The envisioned architecture for PMU data exchange in the US power grid, also referred to as the North American Synchrophasor Initiative Network, or NASPInet \cite{naspinet}, involves PMUs located at substations of different utility companies sending their measurements to controllers at remote generators through a wide-area communication network such as the Internet. To reduce the cost of this communication, sparsity promotion for WAC was studied in \cite{dorjovchebulTPS14,lin2013design}. However, these designs employed social centralized implementation. In contrast, our approach is to design distributed controllers for WAC that are efficient with respect to communication costs. Moreover, we address the question posed in \cite{naspinet} on financing the communication network among the power companies. While the envisioned NASPInet has many functions beyond WAC, fair allocation of network costs associated with delivery of feedback data is necessary to develop its overall pricing scheme.


We model a power transmission system using (\ref{short:eq},\ref{partition:eq}--\ref{K^i:eq}) and as in Fig.\ref{comm:fig}, where a node represents a {\it generator} while an agent (player) represents the operating territory ({\it area}) owned by a utility company. Following \cite{dorjovchebulTPS14,lianensuring}, we express WAC as an LQR problem for minimizing the closed-loop energy of the system states, which equivalently translates into reducing the oscillations in their dynamic response. The LQR objective in (\ref{J:eq}) aims to damp the power oscillations captured by the small-signal changes around the nominal values while spending a reasonable amount of control effort \cite{dorjovchebulTPS14}, where the matrices $\boldsymbol Q$ and $\boldsymbol R$ in (\ref{J:eq}) are chosen to reduce the energies of the state and control vectors, respectively. Therefore, in the rest of this paper we will refer to the objective (e.g., in (\ref{J:eq}), (\ref{Ji:eq}), Algorithm \ref{alg3:alg}, etc.) as {\it energy} and to the payoffs $v_{\mathrm{soc}}(s)$ and $v_i(s)$ in Algorithm \ref{alg3:alg} (Step 4 (1 and 2)) as {\it energy savings}.

Typically a $3^{\mathrm{rd}}$-order model of synchronous generators, including two swing states and one excitation state (for details, please see \cite{pramod}) suffices for solving most WAC problems since the goal is to influence the electro-mechanical dynamics of these generators. Note that the $3^{\mathrm{rd}}$-order model indicates $m_j = 3$ and $p_j = 1, \forall j = 1,...,n$ in the general state-space model ({\ref{short:eq}}). We consider a power system with $n$ synchronous generators, divided into $r$ non-overlapping areas, and define the states of the $j^{th}$ generator accordingly as
\begin{equation}
{\mathcal X}_j^{\mathrm{T}}(t) = \left(\Delta \delta_j(t), \Delta \omega_j(t), \Delta E_j(t)\right),
\end{equation}
\noindent where the elements represent small-signal changes in phase angle, frequency, and excitation voltage, respectively. If higher-order models are considered, then the last term can be simply replaced by a vector $\boldsymbol{x}_j^{-}$ that collects all states except for the phase and the frequency. We assume full state availability, which can be achieved by placing PMUs at every generator bus, or by running a prior state estimation loop. We assume the matrix $\boldsymbol{D}$ in (\ref{short:eq}) to be an indicator matrix with all elements zero except for the one corresponding to the acceleration equation of the generator at which the fault $w(t)$ happens.


Next, we describe the choice of matrices in the social (\ref{J:eq}) and the selfish (\ref{Ji:eq}) energy optimization. We set $\boldsymbol{R}$ in (\ref{J:eq}) and ${\boldsymbol R}_i$ (\ref{Ji:eq}) as the identity matrix as to achieve the same weight for the energy of every control input. The matrix ${\boldsymbol Q}$ in (\ref{J:eq}) is chosen so that all generators arrive at a consensus in their small-signal changes in phase angles and frequencies, as dictated by the physical topology of the network \cite{lianensuring,dorjovchebulTPS14}. Considering the small-signal-model in Kron-reduced form \cite{pramod}, for the $3^{\mathrm{rd}}$ order model, $\boldsymbol{Q}$ in (\ref{J:eq}) is determined from (\ref{Qwac:eq})
\begin{align}
&E_{\mathrm{states}} =  \begin{bmatrix}\Delta {\boldsymbol \delta}\\ \Delta {\boldsymbol \omega} \\ \Delta {\boldsymbol E}\end{bmatrix}^T
\underbrace{\begin{bmatrix}\bar {\mathcal L} & ~ &~\\
~& \bar {\mathcal L} & ~\\
~&~& {\boldsymbol I}\end{bmatrix} }_{{\boldsymbol Q}'}
\begin{bmatrix}\Delta {\boldsymbol \delta}\\ \Delta {\boldsymbol \omega} \\ \Delta {\boldsymbol E}\end{bmatrix} \nonumber\\ 
&= {\boldsymbol x}^T({\mathcal P}^T{\boldsymbol Q'}{\mathcal P}){\boldsymbol x} = {\boldsymbol x}^T{\boldsymbol Q}{\boldsymbol x}\nonumber \\
&=\sum_{k=1}^{n}{\sum_{j=k+1}^{n}{[(\Delta \delta_j - \Delta \delta_k)^2 + (\Delta \omega_j -\Delta \omega_k)^2]}} + \sum_{j=1}^{n}{\Delta E_j^2}, 
\label{Qwac:eq}
\end{align}

\noindent where $\mathcal P$ is a permutation matrix that rearranges the state vector $\boldsymbol x$ in (\ref{short:eq}) by stacking all the angles first, then all the frequencies, then the excitation voltages. For the general case (\ref{short:eq}), the $\Delta {\boldsymbol E}^{\mathrm{T}}\Delta {\boldsymbol E}$ term in (\ref{Qwac:eq}) is replaced by $({\boldsymbol x}^-)^{\mathrm{T}}{\boldsymbol x}^-$, where the latter is obtained by stacking the terms $\boldsymbol{x}_j^{-}$. The matrix $\bar {\mathcal L} = n{\boldsymbol I}-{\boldsymbol 1}_n \cdot{\boldsymbol 1}_n^T$ \cite{Lian:aa}, where $\boldsymbol{1}_n\in \mathbb{R}^{n \times 1}$ is the column vector of all ones, and $\boldsymbol{I}$ is the identity matrix.

To quantify the $i^{\mathrm{th}}$ player's selfish objective (\ref{Ji:eq}), we define two quantities, namely, {\it intra-area energy} and {\it inter-area energy}, from the perspective of the $i^{\mathrm{th}}$ player as follows:
\begin{gather}
\resizebox{\eqratio\textwidth}{!}{
$\begin{aligned}
\label{Ei_intra:eq}
E_i^{\mathrm{intra}}({\boldsymbol{x}}):=\sum_{k\in s_i}\sum_{j\in s_i \atop j>k}{\big[(\Delta \delta_k-\Delta \delta_j)^2 +(\Delta \omega_k-\Delta \omega_j)^2\big]} + \sum_{k\in s_i}{\Delta E_k^2 }\\
\end{aligned}$}\raisetag{1\baselineskip}
\end{gather} 
\vspace{-9pt}
\begin{gather}
\resizebox{\eqratio\textwidth}{!}{
$\begin{aligned}
E_i^{\mathrm{inter}}(\boldsymbol{x}):=
\frac{1}{2}\sum_{k\in s_i}\sum_{j=1,...,n,\atop j \notin s_i}{\big[(\Delta \delta_k-\Delta \delta_j)^2 + (\Delta \omega_k-\Delta \omega_j)^2\big]}
\label{Ei_inter:eq}
\end{aligned}$}\raisetag{1\baselineskip}
\end{gather} 

\noindent
It can be seen that the intra-area energy of area $i$ is designed for the consensus in the phase angle and frequency states of the generators in area $i$. The inter-area energy is modeled by collecting the power transfer terms associated with a generator in area $i$ and a generator in another area, and attributing 1/2 of this energy to area $i$.
\noindent The total {\it state energy} associated with area $i\in\{1,...,r\}$ is 
\begin{equation}
{\boldsymbol x}^T{\boldsymbol Q}_i{\boldsymbol x} = E_i^{\mathrm{intra}}(\boldsymbol{x}) + E_i^{\mathrm{inter}}(\boldsymbol{x}),
\label{Qi_cross:eq}
\end{equation}

\noindent where $\boldsymbol{Q}_i \geq 0$, since (\ref{Qi_cross:eq}) is the quadratic form of the states. Detailed derivations of ${\boldsymbol Q}$ and ${\boldsymbol Q}_i$ can be found in \cite{Lian:aa}. It is easy to show that the resulting LQR objectives (\ref{J:eq}) and (\ref{Ji:eq}) satisfy (\ref{sumQ:eq}), thus assuring successful cooperation among the power companies. We illustrate this fact in the next section by numerical simulation of a 50-bus power system model.


\section{Numerical Results and Performance Analysis for the Australian Power System Model}

\label{num:sec}

\begin{figure}[!t]
  	\centering
  	\subfigure[Line diagram of the Australian 50-bus system.]{\includegraphics[width=\auswidthb]{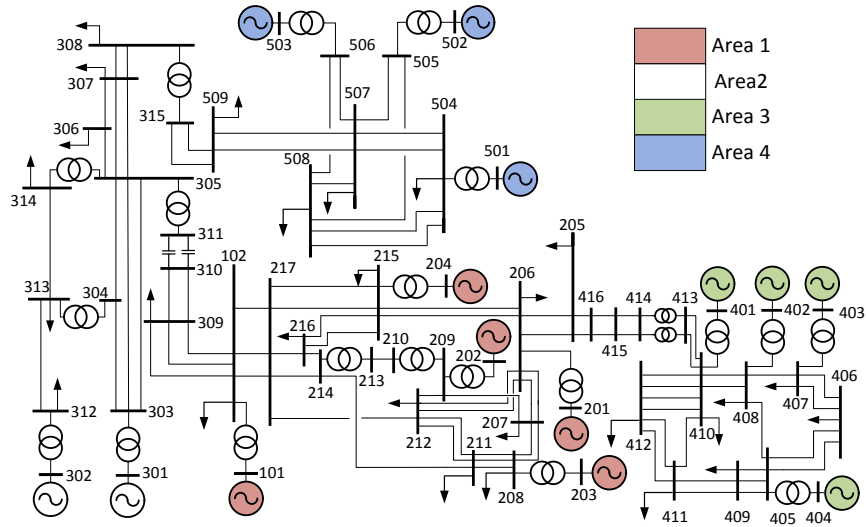}}
  	\subfigure[Simplified Australian power system with feedback links.]{\includegraphics[width=\auswidth]{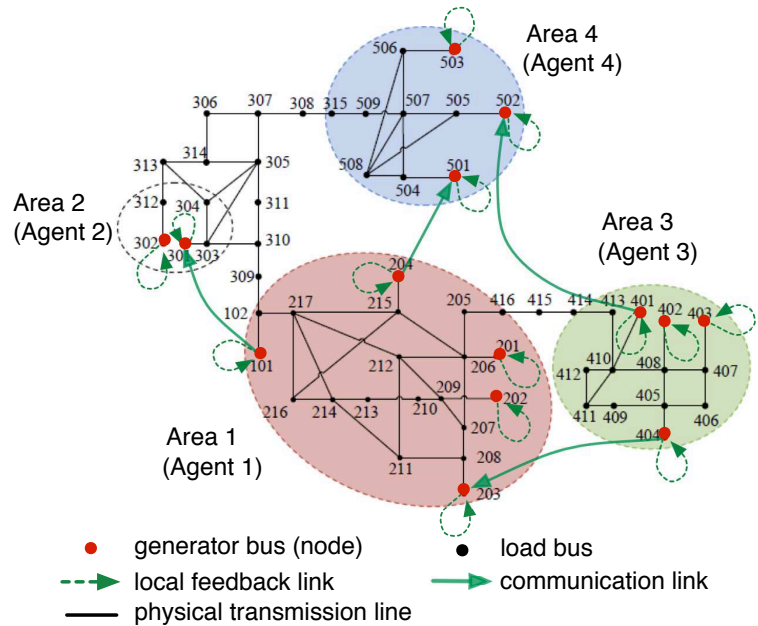}}
  	\caption{A simplified 50-bus representation of the southeast Australian power system \cite{Gibbard:2010aa}.}
  	\label{aus:fig}
  \end{figure}

We validate our results using a 50-bus Australian power system model shown in Fig.\ref{aus:fig}(a), which consists of 14 synchronous generators, divided into 4 coherent areas, and is a reasonably accurate representation of the power grid in south-eastern Australia \cite{Gibbard:2010aa}. The area distribution is shown in different colors in Fig.\ref{aus:fig}(b), with the red dots denoting generator buses. Generators 1 to 5 belong to area 1, 6 and 7 -- to area 2, 8 to 11 -- to area 3, and 12 to 14 -- to area 4. Each generator is modeled by up to 17 states, namely the generator phase angle, the generator speed, direct and quadrature axis components of the internal voltage of the generator, direct and quadrature axis components of the internal flux of the generator, the field excitation voltage, three states contributed by the automatic voltage regulator (AVR), three states contributed by the power system stabilizer (PSS), one state contributed by the stabilizing transformer, and finally three states contributed by supporting induction generators. The small-signal linearized model is extracted using the MATLAB PST toolbox \cite{Chow:1992aa}. However, since we are primarily interested in the electro-mechanical states, we perform an initial round of model reduction using singular perturbation and thereby eliminate the non-electromechanical states with very low participation in the swing dynamics. The exact expressions of the model matrices are not included in the paper for brevity and are provided in \cite{Lian:aa}. The small-signal model is excited by impulsive disturbance inputs entering through the acceleration equation of the generators, and the proposed LQR controller is actuated through the field excitation voltages, using state feedback from all generators. Fig.\ref{aus:fig}(b) also illustrates the communication and local feedback links between the generators. 

\subsection{Global Energies}
\label{alg_pfm:sec}

\begin{figure}[!t]
\centering 
\includegraphics[width=\curvewidth]{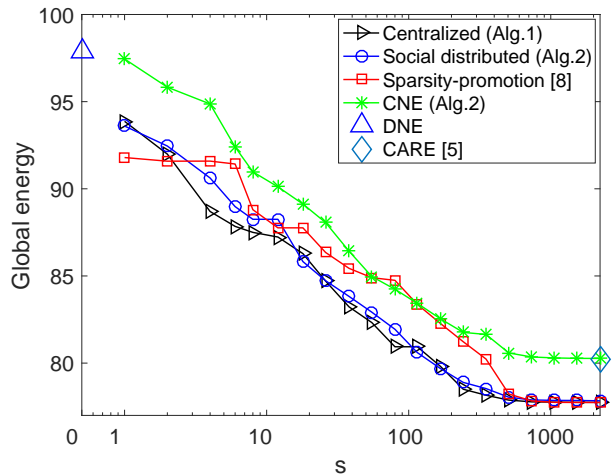}
\caption{Global energies of social optimization and noncooperative games vs. communication cost constraint.}
\label{J:fig}
\end{figure}

Fig.\ref{J:fig} shows the global closed-loop energies for the 50-bus system with respect to the communication cost constraints for the various centralized and distributed algorithms discussed in the previous sections. Global energies of the centralized optimization (\ref{sparse_cent:eq}) using Algorithm \ref{cent:alg}, the coupled noncooperative game (CNE) in Step 3 of Algorithm \ref{alg3:alg}, the social distributed optimization (\ref{NE_J:eq}) using Algorithm 2, and of the decoupled game (DNE) in Step 2 of Algorithm \ref{alg3:alg} are included. We also show the global energy of the iterative dense-feedback method that solves CARE \cite{Mukaidani2006} (note that our system satisfies the weakly coupled condition in \cite{Mukaidani2006}) and of the centralized sparsity-promoting ADMM method \cite{dorjovchebulTPS14}, modified to satisfy the sparsity constraint. Since the energy of the method in \cite{dorjovchebulTPS14} is nondecreasing with the sparsity parameter $\gamma$, a bisection search on $\gamma$ yields the smallest value of $\gamma$ for which the off-diagonal cardinality of the feedback matrix produced by the ADMM algorithm satisfies the constraint. The choice of the $l_1$-metric weights for this ADMM implementation is described in \cite{lianensuring}.

The figure shows that the global closed-loop energies resulting from the social optimization using Algorithm {\ref{cent:alg}} or {\ref{graspNE:alg}} are smaller than those of the noncooperative game (CNE using Algorithm \ref{graspNE:alg}) for any value of $s$. This testifies to the fact that the dynamic performance of the grid improves when companies cooperate with each other to jointly design the controller, which constitutes a sufficient condition for the cooperation to form according to ({\ref{eqB:eq}}). We also observe that the global CNE objective (Algorithm \ref{alg3:alg}, Step 3(3)) tends to those of the DNE (Algorithm \ref{alg3:alg}, Step 2(3)) and CARE as $s$ approaches 0 and its largest value $s=2223$, respectively. The former corresponds to the decentralized case \cite{lianensuring} while the latter requires dense communication \cite{Mukaidani2006}.

Moreover, the energy $J_{\mathrm{soc}}$ of the distributed social optimization (using Algorithm 2) closely approximates that of the centralized Algorithm 1 for most of the $s$-range. We also observe that the closed-loop energy of social optimization using Algorithms {\ref{cent:alg}} or {\ref{graspNE:alg}} is smaller than that using the sparsity-promoting algorithm in a moderately sparse region, thus providing better reduction of both intra- and inter-area oscillations when given an exact communication cost constraint.

While the global energies of all sparsity-constrained methods are theoretically nonincreasing with $s$, the social algorithms might occasionally produce a larger energy as $s$ increases. This happens when an algorithm converges to a local minimum since the optimization objective (\ref{J:eq}) is locally, but not necessarily globally, convex in $\boldsymbol{K}$ \cite{rautert1997computational}. If the algorithm results in $J_{\mathrm{soc}}(s_2){>} J_{\mathrm{soc}}(s_1)$ for $s_2{>}s_1$, we choose a suboptimal solution $J_{\mathrm{soc}}(s_2)\triangleq J_{\mathrm{soc}}(s_1)$ (see Algorithm \ref{alg3:alg}, Step 1), which produces a nonincreasing $J_{\mathrm{soc}}(s)$. Finally, note that the global objectives (energies) of all algorithms saturate to the same asymptotic value when $s$ exceeds 740, implying that the communication cost can be reduced by a factor of 3 relative to the cost of the dense LQR network without compromising the control performance. Thus, in the next figure, we show results only for small and moderate values of $s$, where the energies vary significantly.

\subsection{Selfish Energies and Cost Allocation}
\label{cost_alloc:sec}

\begin{figure}[!t]
\centering 
\includegraphics[width=\figfourwidth]{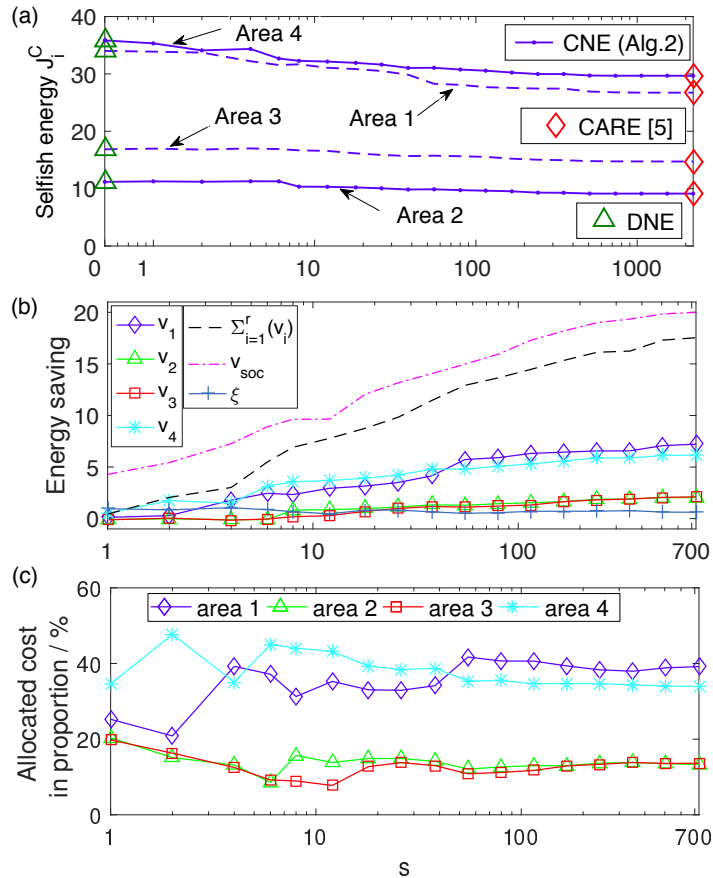}
\caption{Selfish energies (objectives), payoffs and cost allocation in Algorithm \ref{alg3:alg} vs. communication cost constraint $s$. (a) Selfish energies of noncooperative games for areas 1--4. (b) The energy savings (payoffs) $v_i(s)$ and $v_{\mathrm{soc}}(s)$ and payoff increase $\xi(s)$. (c) Proportional cost allocation.}
\label{payoff:fig}
\end{figure}

In Fig.\ref{payoff:fig}, we show performance of {\it noncooperative} games, as well as payoffs and costs in Algorithm \ref{alg3:alg}. Fig.\ref{payoff:fig}(a) shows the individual energy objectives of the four areas at CNE (Step 3(2)). In Fig.\ref{payoff:fig}(b), the selfish energy savings $v_i(s)$ (Step 4(2)) for each company, their sum, the social energy savings $v_{\mathrm{soc}}(s)$ (Step 4(1)), and the payoff increase $\xi(s)$ (\ref{delta:eq}) are illustrated. We observe that bargaining is successful (see (\ref{eqB:eq})), and there is modest payoff increase due to cooperation. 

Note that while the overall global CNE objective $\tilde J^{\mathrm{C}}(s)$ (Step 3(3)) is theoretically nonincreasing with $s$, this is not necessarily true for individual selfish energies $J_i^{\mathrm{C}}(s)$ in Step 3(2), which means that increasing the overall communication budget might degrade some areas' energies when they act noncooperatively. This phenomenon can result in decreasing and possibly negative payoffs $v_i(s)$ and $\alpha_i(s)$ over some regions of the constraint $s$, which might dissatisfy the affected players and would require a revision of the proposed cost allocation method in Step 4(4) of Algorithm \ref{alg3:alg}. See, e.g., slightly increasing selfish objective $J_3^{\mathrm{C}}(s)$ and negative selfish payoff $v_3(s)$ of area 3 in Fig.{\ref{payoff:fig}} (a) and (b), respectively, for small values of $s$. Nevertheless, the allocated payoffs $\alpha_i(s)$ in Algorithm \ref{alg3:alg} (Step 4(3)) are nonnegative for all companies for the Australian power system scenario due to sufficient cooperation gain $\xi(s)$. A slightly modified payoff computation method that guarantees nondecreasing payoffs is presented in Appendix \ref{B:appendix}.

Fig.\ref{payoff:fig} demonstrates significant disparity in the selfish energies $J_i^{\mathrm{C}}(s)$, the $v_i(s)$ values, and the allocated proportional costs $C_i(s)$ among the areas, which is due to the grid topology. For example, large selfish energy and allocated cost of area 1 can be explained by its large number of generators. However, area size is not the only indicator, e.g., area 4 has fewer generators than area 3, but much larger selfish energy $J_4^{\mathrm{C}}(s)$, energy savings $v_4(s)$, and allocated cost $\alpha_4(s)$, which even exceed those of area 1 for smaller values of $s$. In summary, areas 1 and 4 pay a much greater share of the overall network cost than areas 2 and 3 due to the former areas' greater needs for feedback and cooperation, which is consistent with relatively steep decline of their selfish energies with $s$ in Fig.\ref{payoff:fig}(a) and with the fact that in the social optimization, links among the generators in these areas are the first to be added as $s$ increases. Thus, these links are the most valuable for achieving energy reduction when using WAC \cite{dorjovchebulTPS14}.

\subsection{Algorithm Convergence and Implementation Issues}
\label{convergence:sec}

First, we focus on the numerical properties of Algorithms 1 and 2. Since the LQR objective does not satisfy the Stable Restricted Hessian condition \cite{bahmani2013greedy}, convergence of these algorithms is not assured in general. However, if Step 2 of Algorithm 1 converges and yields a stabilizing feedback matrix ${\boldsymbol K}$, then Step 3 will also converge due to the local convexity of $J({\boldsymbol K})$. At convergence, Step 3 produces a feedback matrix $\hat{\boldsymbol K}$ that satisfies the basic feasibility property $\nabla_{\boldsymbol K}J(\boldsymbol K)|_{\mathrm{supp}(\hat{\boldsymbol K})}(\boldsymbol K=\hat{\boldsymbol K}) = 0$, which is a weak necessary condition for the optimality of problem (\ref{sparse_cent:eq}) \cite{beck2013sparsity}. Similar arguments demonstrate that Algorithm 2 produces a feedback matrix that satisfies the basic feasibility property of each individual minimization in the CNE problem (\ref{NE_sparseK:eq}) (or (\ref{NE_J:eq}) for social optimization). Note that the decoupled game in Step 2 of Algorithm {\ref{alg3:alg}} represents an unconstrained optimization with respect to each player's strategy, which is given by its local feedback matrix \cite{lianensuring}. This game is implemented using the polishing Step 3 of Algorithm {\ref{graspNE:alg}}. When this implementation converges, the resulting equilibrium point is a local NE \cite{ratliff2013characterization} due to local convexity of the LQR objective \cite{rautert1997computational}, i.e., the strategy of each player is a local optimum given other players' strategies. From these observations, we conclude that convergence properties of proposed algorithms resemble those of the ADMM-based methods \cite{dorjovchebulTPS14}. In both cases, while theoretical guarantees are not always feasible, extensive numerical experience demonstrates that the algorithms converge and provide desirable minimization solutions over a range of sparsity parameters. We found that the proposed sparsity-constrained algorithms also converge and exhibit similar performance and complexity trends to those shown in this paper for the New England power system model used in \cite{dorjovchebulTPS14}.

Next, we describe the algorithm implementation details for the results shown in Fig.\ref{J:fig}--\ref{payoff:fig}. We found that the proposed algorithms can converge to different stabilizing feedback matrices given different initial settings. Moreover, for $s>0$, the energies of these solutions can differ significantly. However, we found that the energies of different equilibria were very similar to each other for the decoupled game implemented using Algorithm \ref{alg3:alg} (Step 2). We have employed $\epsilon_{\mathrm{abs}}=\epsilon_{\mathrm{rel}}=10^{-4}$ and $\epsilon_2= 10^{-4}$ and $10^{-3}$ in Algorithms 1 and 2, respectively, to achieve comparable performance for distributed and centralized social optimizations. For $s=1$, both algorithms were initialized with a stabilizing matrix ${\boldsymbol K}_0$ obtained by preserving the block-diagonal entries of the dense LQR feedback matrix that optimizes (\ref{J:eq}) and setting other entries to zero. For larger $s$, $\boldsymbol{K}_0$ was chosen as the optimized feedback matrix obtained in a previous computation for a smaller value of $s$. We found that this initialization produced the lowest energies over the entire $s$-range. If a stable block-diagonal matrix cannot be found, ${\boldsymbol K}_0$ can be obtained using the sparsity-promotion algorithm \cite{dorjovchebulTPS14} with the largest $\gamma$ that produces a stabilizing feedback matrix (Similarly, stabilizing feedback might not exist for small values of the constraint $s$). Moreover, Algorithm 2 had the best performance when the initial link settings $s_i$ were chosen proportionally to the number of nodes $n_i$ in (\ref{partition:eq}).

Finally, we found that the computational load of the polishing step dominates the overall runtime for both algorithms, and the Newton step using the CG method (see Table \ref{notation:tb}, last entry) is the most computation-intensive operation, which has polynomial complexity in $s$ and the number of states \cite{lin2013design}. In our experiments, all algorithms in Fig.\ref{J:fig} converged in less than $10^3$ seconds for any value of $s$ although this did not include the bisection search time for the modified ADMM method \cite{dorjovchebulTPS14}, which is very computation-intensive. Moreover, the distributed social implementation using Algorithm 2 converged much faster than the centralized method (Algorithm 1)\footnote{The experiments are run using MATLAB on a MacBook Pro with Yosemite operating system, 2.6 GHz Intel core i5 processor, and 8 GB 1600 MHz DDR3 memory.}.

\section{Conclusion}

LQR optimization under the communication-cost constraint was investigated for multi-agent dynamic systems with linear static state feedback. First, a communication-cost-constrained centralized social optimization algorithm was developed. Second, distributed game-theoretic algorithms were investigated for both selfish and social optimization under the sparsity constraint. Finally, cooperative NFG theory was employed to allocate the costs of the communication infrastructure in a multi-agent dynamic system. Using a 50-bus power system model divided into 4 areas, we demonstrated convergence of proposed algorithms and desirable performance and complexity features of distributed methods over the range of the sparsity constraint, thus providing a trade-off between the communication cost and the control performance. Furthermore, we discussed the relationship between the proportionally allocated costs and power companies' needs for feedback and cooperation, and showed that the proposed cost allocation algorithm is rooted in the physical topology of the power grid. Our current research focuses on applying proposed algorithms to systems with uncertainty and communication delays. Moreover, we plan to extend the concepts of this paper to diverse multi-agent dynamic system applications, robust optimization objectives, output feedback, and different communication network architectures.

\section*{Acknowledgment}
The authors would like to thank the editors and reviewers for their valuable comments, which improved the paper significantly.




%


\bibliographystyle{IEEEtran}
\bibliography{ref}

\begin{thebibliography}{10}
\providecommand{\url}[1]{#1}
\csname url@samestyle\endcsname
\providecommand{\newblock}{\relax}
\providecommand{\bibinfo}[2]{#2}
\providecommand{\BIBentrySTDinterwordspacing}{\spaceskip=0pt\relax}
\providecommand{\BIBentryALTinterwordstretchfactor}{4}
\providecommand{\BIBentryALTinterwordspacing}{\spaceskip=\fontdimen2\font plus
\BIBentryALTinterwordstretchfactor\fontdimen3\font minus
  \fontdimen4\font\relax}
\providecommand{\BIBforeignlanguage}[2]{{%
\expandafter\ifx\csname l@#1\endcsname\relax
\typeout{** WARNING: IEEEtran.bst: No hyphenation pattern has been}%
\typeout{** loaded for the language `#1'. Using the pattern for}%
\typeout{** the default language instead.}%
\else
\language=\csname l@#1\endcsname
\fi
#2}}
\providecommand{\BIBdecl}{\relax}
\BIBdecl

\bibitem{sztipanovits2012toward}
J.~Sztipanovits, X.~Koutsoukos, G.~Karsai, N.~Kottenstette, P.~Antsaklis,
  V.~Gupta, B.~Goodwine, J.~Baras, and S.~Wang, ``Toward a science of
  cyber--physical system integration,'' \emph{Proceedings of the IEEE}, vol.
  100, no.~1, pp. 29--44, 2012.

\bibitem{basar85}
T.~Ba\c{s}ar and G.~J. Olster, \emph{Dynamic noncooperative game theory}.\hskip
  1em plus 0.5em minus 0.4em\relax SIAM, 1995, vol. 200.

\bibitem{lewis1995optimal}
F.~L. Lewis and V.~L. Syrmos, \emph{Optimal control}.\hskip 1em plus 0.5em
  minus 0.4em\relax John Wiley \& Sons, 1995.

\bibitem{LUKES197196}
D.~Lukes and D.~Russell, ``A global theory for linear-quadratic differential
  games,'' \emph{Journal of Mathematical Analysis and Applications}, vol.~33,
  no.~1, pp. 96--123, 1971.

\bibitem{Mukaidani2006}
H.~Mukaidani, ``A numerical analysis of the {N}ash strategy for weakly coupled
  large-scale systems,'' \emph{IEEE Transactions on Automatic Control},
  vol.~51, no.~8, pp. 1371--1377, Aug. 2006.

\bibitem{bahmani2013greedy}
S.~Bahmani, B.~Raj, and P.~T. Boufounos, ``Greedy sparsity-constrained
  optimization,'' \emph{The Journal of Machine Learning Research}, vol.~14,
  no.~1, pp. 807--841, 2013.

\bibitem{beck2013sparsity}
A.~Beck and Y.~C. Eldar, ``Sparsity constrained nonlinear optimization:
  Optimality conditions and algorithms,'' \emph{SIAM Journal on Optimization},
  vol.~23, no.~3, pp. 1480--1509, 2013.

\bibitem{dorjovchebulTPS14}
F.~D\"{o}rfler, M.~R. Jovanovi\'{c}, M.~Chertkov, and F.~Bullo, ``Sparse and
  optimal wide-area damping control in power networks,'' in \emph{American
  Control Conference (ACC)}, 2013, pp. 4289--4294.

\bibitem{lin2013design}
F.~Lin, M.~Fardad, and M.~R. Jovanovic, ``Design of optimal sparse feedback
  gains via the {A}lternating {D}irection {M}ethod of {M}ultipliers,''
  \emph{IEEE Transactions on Automatic Control}, vol.~58, no.~9, pp.
  2426--2431, 2013.

\bibitem{lamperski2015optimal}
A.~Lamperski and L.~Lessard, ``Optimal decentralized state-feedback control
  with sparsity and delays,'' \emph{Automatica}, vol.~58, pp. 143--151, 2015.

\bibitem{mota2013d}
J.~F. Mota, J.~M. Xavier, P.~M. Aguiar, and M.~Puschel, ``{D-ADMM}: {A}
  communication-efficient distributed algorithm for separable optimization,''
  \emph{IEEE Transactions on Signal Processing}, vol.~61, no.~10, pp.
  2718--2723, 2013.

\bibitem{fardad2009optimal}
M.~Fardad, F.~Lin, and M.~R. Jovanovi{\'c}, ``On the optimal design of
  structured feedback gains for interconnected systems,'' in \emph{Proceedings
  of the 48th IEEE Conference on Decision and Control}, 2009, pp. 978--983.

\bibitem{ratliff2013characterization}
L.~J. Ratliff, S.~A. Burden, and S.~S. Sastry, ``Characterization and
  computation of local {N}ash equilibria in continuous games,'' in \emph{51st
  Annual Allerton Conference on Communication, Control, and Computing
  (Allerton)}.\hskip 1em plus 0.5em minus 0.4em\relax IEEE, 2013, pp. 917--924.

\bibitem{li2013designing}
N.~Li and J.~R. Marden, ``Designing games for distributed optimization,''
  \emph{IEEE Journal of Selected Topics in Signal Processing}, vol.~7, no.~2,
  pp. 230--242, 2013.

\bibitem{van2005models}
A.~van~den Nouweland, ``Models of network formation in cooperative games,''
  \emph{Group formation in economics}, pp. 58--88, 2005.

\bibitem{Avrachenkov2015265}
K.~Avrachenkov, J.~Elias, F.~Martignon, G.~Neglia, and L.~Petrosyan,
  ``Cooperative network design: A {N}ash bargaining solution approach,''
  \emph{Computer Networks}, 2015.

\bibitem{kawamori2016nash}
T.~Kawamori and T.~Miyakawa, ``{N}ash bargaining solution under
  externalities,'' \emph{Mathematical Social Sciences}, vol.~84, pp. 1--7,
  2016.

\bibitem{Lian:2014aa}
F.~Lian, A.~Duel-Hallen, and A.~Chakrabortty, ``Cost allocation strategies for
  wide-area control of power systems using {N}ash bargaining solution,'' in
  \emph{IEEE 53rd Annual Conference on Decision and Control}, Dec 2014, pp.
  1701--1706.

\bibitem{lianensuring}
------, ``Ensuring economic fairness in wide-area control for power systems via
  game theory,'' in \emph{American Control Conference}, 2016.

\bibitem{naspinet}
P.~T. Myrda and K.~Koellner, ``{NASPI}net -- the {I}nternet for
  synchrophasors,'' in \emph{IEEE 43rd Hawaii International Conference on
  System Sciences (HICSS)}, 2010, pp. 1--6.

\bibitem{pramod}
A.~Chakrabortty and P.~Khargonekar, ``Introduction to wide-area control of
  power systems,'' in \emph{American Control Conference, DC}, 2013.

\bibitem{deng2012communication}
Y.~Deng, H.~Lin, A.~G. Phadke, S.~Shukla, J.~S. Thorp, and L.~Mili,
  ``Communication network modeling and simulation for wide area measurement
  applications,'' in \emph{2012 IEEE PES Innovative Smart Grid Technologies
  (ISGT)}.\hskip 1em plus 0.5em minus 0.4em\relax IEEE, 2012, pp. 1--6.

\bibitem{chenine2009survey}
M.~Chenine, K.~Zhu, and L.~Nordstrom, ``Survey on priorities and communication
  requirements for {PMU}-based applications in the {N}ordic region,'' in
  \emph{2009 IEEE Bucharest PowerTech}.\hskip 1em plus 0.5em minus 0.4em\relax
  IEEE, 2009, pp. 1--8.

\bibitem{pthorp}
A.~G. Phadke and J.~S. Thorp, \emph{Synchronized phasor measurements and their
  applications}.\hskip 1em plus 0.5em minus 0.4em\relax Springer Science \&
  Business Media, 2008.

\bibitem{rautert1997computational}
T.~Rautert and E.~W. Sachs, ``Computational design of optimal output feedback
  controllers,'' \emph{SIAM Journal on Optimization}, vol.~7, no.~3, pp.
  837--852, 1997.

\bibitem{Boyd:2004aa}
S.~Boyd and L.~Vandenberghe, \emph{Convex optimization}.\hskip 1em plus 0.5em
  minus 0.4em\relax Cambridge university press, 2004.

\bibitem{pang2011nonconvex}
J.-S. Pang and G.~Scutari, ``Nonconvex games with side constraints,''
  \emph{SIAM Journal on Optimization}, vol.~21, no.~4, pp. 1491--1522, 2011.

\bibitem{Peters2008}
H.~Peters, \emph{Cooperative Games with Transferable Utility}.\hskip 1em plus
  0.5em minus 0.4em\relax Springer Berlin Heidelberg, 2008, pp. 121--131.

\bibitem{saad2009distributed}
W.~Saad, Z.~Han, M.~Debbah, and A.~Hj{\o}rungnes, ``A distributed coalition
  formation framework for fair user cooperation in wireless networks,''
  \emph{IEEE Transactions on Wireless Communications}, vol.~8, no.~9, pp.
  4580--4593, 2009.

\bibitem{myerson1980conference}
R.~B. Myerson, ``Conference structures and fair allocation rules,''
  \emph{International Journal of Game Theory}, vol.~9, no.~3, pp. 169--182,
  1980.

\bibitem{Hafalir2007242}
I.~E. Hafalir, ``Efficiency in coalition games with externalities,''
  \emph{Games and Economic Behavior}, vol.~61, no.~2, pp. 242 -- 258, 2007.

\bibitem{Lian:aa}
\BIBentryALTinterwordspacing
F.~Lian, ``Supplementary materials for `{G}ame-theoretic multi-agent control
  and network cost allocation under communication constraints', submitted to
  \textit{{IEEE} {JSAC} Special Issue on Game Theory for Networks}, 2016.''
  [Online]. Available:
  \url{http://www4.ncsu.edu/~flian2/jsac2016_supplement.html}
\BIBentrySTDinterwordspacing

\bibitem{Gibbard:2010aa}
M.~Gibbard and D.~Vowles, ``Simplified 14-generator model of the se australian
  power system,'' \emph{Technical Report, The University of Adelaide, South
  Australia}, pp. 1--45, 2010.

\bibitem{Chow:1992aa}
J.~Chow and K.~Cheung, ``A toolbox for power system dynamics and control
  engineering education and research,'' \emph{IEEE Trans. Power Syst.}, vol.~7,
  no.~4, pp. 1559--1564, Nov 1992.

\end{thebibliography}


\begin{thebibliography}{}

\bibitem[Saad, 2010]{Saad:2010aa}
W. Saad, ``Coalitional Game Theory for Distributed Cooperation in Next Generation Wireless Networks," Ph.D. dissertation, University of Oslo, 2010.

\end{thebibliography}

%








\appendices

\section{Definition of Nondecreasing Selfish Payoffs for Algorithm \ref{alg3:alg}}

\label{B:appendix}
In Section \ref{cost_alloc:sec}, we comment that it is possible to define nondecreasing selfish payoffs in Step 4(2) of Algorithm \ref{alg3:alg}. Given a cost constraint $s$, such alternative definition is 
\begin{equation}
v_i^*(s) = J_i^{\mathrm{D}}-\underbrace{\mathrm{min}(\{J_i^{\mathrm{C}}(s')| s'\leq s\})}_{J_i^*(s)}, i=1,...,r
\label{vi_star:eq}
\end{equation}

\noindent which is the maximum objective improvement an agent $i$ can obtain by searching over the set of its selfish objectives $J_i^{\mathrm{C}}(s')$ with constraints $s'$ that do not exceed $s$. If there exists some $s' < s$ such that $J_i^{\mathrm{C}}(s') < J_i^{\mathrm{C}}(s)$, the agent $i$ might argue that the smaller selfish objective $J_i^{\mathrm{C}}(s')$, not $J_i^{\mathrm{C}}(s)$, should be used to compute its payoff since it also satisfies the constraint ($s'<s$). It is easy to show that $v_i^*(s)$ in (\ref{vi_star:eq}) is non-decreasing with $s$ and $v_i^*(s)\geq0, \forall s\geq 0$. 


 Note that the disagreement point (\ref{vi_star:eq}) is hypothetical in a sense that a communication network with the energies $J_i^*(s)$ in (\ref{vi_star:eq}) might not be feasible (different agents might have different $s'$ values in (\ref{vi_star:eq}) for a fixed constraint $s$). However, noncompatible selfish payoffs are often employed in the literature to reflect the player's subjective preferences and are not required to represent a feasible scenario \cite{Avrachenkov2015265,kawamori2016nash}. Moreover, successful cooperation ({\ref{eqB:eq}}) is not guaranteed for the payoffs (\ref{vi_star:eq}). However, we found that bargaining was successful for the power system example in Section \ref{num:sec}, and the payoffs ({\ref{vi_star:eq}}) were very similar to the payoffs $v_i(s)$ defined in Algorithm {\ref{alg3:alg}} (Step 4(2)), which were shown in Fig.{\ref{payoff:fig}}(b).

\newpage
\section{Derivations of $\bf Q$ and ${\bf Q}_i$ matrices in Section \ref{WAC:sec}, eq. (\ref{Qwac:eq},\ref{Qi_cross:eq})}

 \subsection{Matrix $\bf{Q}$ in eq. (\ref{Qwac:eq})}

The permutation matrix $\mathcal P$ in eq. (\ref{Qwac:eq}) is
\begin{eqnarray}
\mathcal P =\begin{bmatrix}
\mathcal P_1\\ \hline \mathcal P_2\end{bmatrix}
\end{eqnarray}
\noindent where
\begin{eqnarray}
\mathcal P_2 &=& \mathrm{diag}(\mathcal T_1, \mathcal T_2,..., \mathcal T_n). \\
\mathcal T_i &=& \begin{bmatrix}{\boldsymbol 0}_{(m_i-2)\times 2} &{\boldsymbol I}_{(m_i-2)\times(m_i-2)}\end{bmatrix}\\
\mathcal P_1 &=& (p_{ij})_{2n\times s}
\end{eqnarray}
\noindent and
\begin{eqnarray}
p_{ij} &=& \left\{\begin{array}{cl} \delta_{j,k_i} & , 1\leq i \leq n\\
\delta_{j,k_i+1} & , n+1\leq i \leq 2n \end{array}\right.\nonumber\\
k_i &=& 1+ \sum_{k=1}^{i-1}{m_k}. 
\end{eqnarray}
\noindent Recall that $n$ is the number of nodes in the system, $m_i$ is the number of states belonging to node $i$, $s$ is the total number of states in the network in (\ref{short:eq}), and $\delta_{ij}$ is the Kronecker delta function.

The phase angle terms in (\ref{Qwac:eq}) are given by
\begin{align}
&\sum_{k=1}^{n}\sum_{j=k+1}^{n}{(\Delta \delta_j - \Delta \delta_k)^2} \nonumber\\
&= \frac{1}{2}\sum_{k=1}^{n}\sum_{j=1}^{n}{(\Delta \delta_j - \Delta \delta_k)^2}\nonumber\\
&= \frac{1}{2}\sum_{k=1}^{n}\vert \Delta \delta_k \boldsymbol{1}_{n} -\Delta \boldsymbol{\delta} \vert^2\nonumber\\
&=\frac{1}{2} \sum_{k=1}^{n} n\Delta \delta_k^2 - 2\Delta \delta_i \boldsymbol{1}_n^{\mathrm{T}}\Delta \boldsymbol{\delta} +  \Delta \boldsymbol{\delta}^{\mathrm{T}} \Delta \boldsymbol{\delta} \nonumber\\
&= \frac{1}{2}[n \Delta \boldsymbol{\delta}^{\mathrm{T}} \Delta \boldsymbol{\delta} - 2 \Delta \boldsymbol{\delta}^{\mathrm{T}} \boldsymbol{1}_n \boldsymbol{1}_n^{\mathrm{T}} \Delta\boldsymbol{\delta} 
+ n \Delta \boldsymbol{\delta}^{\mathrm{T}} \Delta \boldsymbol{\delta}] \nonumber\\
&= \Delta \boldsymbol{\delta}^{\mathrm{T}} \left( n\boldsymbol{I}_{n\times n} - \boldsymbol{1}_n \boldsymbol{1}^{\mathrm{T}} \right)\Delta \boldsymbol{\delta} = \Delta \boldsymbol{\delta}^{\mathrm{T}} \bar{\mathcal L} \Delta \boldsymbol{\delta}
\end{align}
\noindent Thus 
\begin{align}
\bar{\mathcal L} = n\boldsymbol{I}_{n\times n} - \boldsymbol{1}_n \boldsymbol{1}^{\mathrm{T}}.
\end{align}

\subsection{Matrix ${\bf Q}_i$ in eq. (\ref{Qi_cross:eq})}

In eq.(24), the intra-area energy for agent $i$
\begin{align}
E_i^{\mathrm{intra}}({\boldsymbol{x}}) &:= \sum_{k\in s_i}\sum_{j\in s_i \atop j>k}{(\Delta \delta_k-\Delta \delta_j)^2 +(\Delta \omega_k-\Delta \omega_j)^2} + \sum_{k\in s_i}{\Delta E_k^2}\nonumber\\
& = \begin{bmatrix}\Delta {\boldsymbol \delta}\\ \Delta {\boldsymbol \omega} \\ \Delta {\boldsymbol E}\end{bmatrix}^T
\begin{bmatrix}\bar {\mathcal L}_i^{\mathrm{intra}} & ~ &~\\
~& \bar {\mathcal L}_i^{\mathrm{intra}} & ~\\
~&~& \mathcal{I}_{i}^{\mathrm{intra}}\end{bmatrix} 
\begin{bmatrix}\Delta {\boldsymbol \delta}\\ \Delta {\boldsymbol \omega} \\ \Delta {\boldsymbol E}\end{bmatrix}
\end{align}

\noindent where $\mathcal I_i^{\mathrm{intra}}$ is a block diagonal matrix with the identity matrix $\boldsymbol{I}_{n_i\times n_i}$ at the $i^{\mathrm{th}}$ diagonal block, and zeros elsewhere.
\begin{align}
\mathcal{I}_i^{\mathrm{intra}} = \mathrm{blkdiag}(\boldsymbol{0}_{n_1\times n_1},..., \underbrace{\boldsymbol{I}_{n_i\times n_i}}_{\mbox{the $i$-th block}},..., \boldsymbol{0}_{n_r\times n_r})
\end{align}
\noindent where $\mathrm{blkdiag}(\boldsymbol{M}_1,...,\boldsymbol{M}_n)$ represents the block-diagonal matrix with matrices $\boldsymbol{M}_1,...,\boldsymbol{M}_n$ on the diagonal blocks. The phase angle terms of (\ref{Ei_intra:eq}) are given by
\begin{align}
\sum_{k\in s_i}\sum_{j\in s_i \atop j>k}{(\Delta \delta_k-\Delta \delta_j)^2} = \Delta {\boldsymbol \delta}^{\mathrm{T}} \bar {\mathcal L}_i^{\mathrm{intra}} \Delta {\boldsymbol \delta}.
\end{align}

\noindent where
\begin{align}
\mathrm{LHS} &= \Delta \boldsymbol{\delta}_i^{\mathrm{T}} \left( n_i \boldsymbol{I}_{n_i\times n_i} - \boldsymbol{1}_{n_i}\boldsymbol{1}_{n_i}^\mathrm{T} \right) \Delta \boldsymbol{\delta}_i \nonumber\\
&= \Delta \boldsymbol{\delta}^{\mathrm{T}} \cdot \mathrm{blkdiag}(\boldsymbol{0}_{n_1\times n_1},..., \underbrace{n_i \boldsymbol{I}_{n_i\times n_i} - \boldsymbol{1}_{n_i}\boldsymbol{1}_{n_i}^\mathrm{T}}_{\mbox{the $i$-th block}},..., \boldsymbol{0}_{n_r\times n_r}) \cdot \Delta \boldsymbol{\delta}\nonumber\\
&= \mathrm{RHS}
\end{align} 
\noindent Thus
\begin{align}
 \bar {\mathcal L}_i^{\mathrm{intra}}= \mathrm{blkdiag}(\boldsymbol{0}_{n_1\times n_1},..., \underbrace{n_i \boldsymbol{I}_{n_i\times n_i} - \boldsymbol{1}_{n_i}\boldsymbol{1}_{n_i}^\mathrm{T}}_{\mbox{the $i$-th block}},..., \boldsymbol{0}_{n_r\times n_r})
\end{align}

In eq.(\ref{Ei_inter:eq}), the inter-area energy for agent $i$
\begin{align}
E_i^{\mathrm{inter}}(\boldsymbol{x})&=\frac{1}{2}\sum_{k\in s_i}\sum_{j=1,...,n,\atop j \notin s_i}{(\Delta \delta_k-\Delta \delta_j)^2 + (\Delta \omega_k-\Delta \omega_j)^2}\nonumber\\
&= \begin{bmatrix}\Delta {\boldsymbol \delta}\\ \Delta {\boldsymbol \omega}\end{bmatrix}^T
\begin{bmatrix}\bar {\mathcal L}_i^{\mathrm{inter}} & ~\\
~& \mathcal{I}_{i}^{\mathrm{inter}}\end{bmatrix}
\begin{bmatrix}\Delta {\boldsymbol \delta}\\ \Delta {\boldsymbol \omega}\end{bmatrix}
\end{align}

\noindent The phase angle terms of (\ref{Ei_inter:eq}) are given by
\begin{align}
\frac{1}{2}\sum_{k\in s_i}\sum_{j=1,...,n,\atop j \notin s_i}{(\Delta \delta_k-\Delta \delta_j)^2} &= \Delta \boldsymbol{\delta}^{\mathrm{T}} \bar {\mathcal L}_i^{\mathrm{inter}}  \Delta \boldsymbol{\delta},
\end{align}
\noindent We express the LHS as
\begin{align}
\mathrm{LHS} &= \frac{1}{2} \sum_{k\in \mathcal{S}_i}\sum_{j=1,j\neq i}^r \vert \Delta \delta_k \boldsymbol{1}_{n_j} - \Delta \boldsymbol{\delta}_j \vert ^2  \nonumber \\
&= \frac{1}{2} \sum_{k\in \mathcal{S}_i}\sum_{j=1,j\neq i}^r \left(\Delta \delta_k \boldsymbol{1}_{n_j} - \Delta \boldsymbol{\delta}_j \right)^{\mathrm{T}} \left(\Delta \delta_k \boldsymbol{1}_{n_j} - \Delta \boldsymbol{\delta}_j \right) \nonumber\\
&= \frac{1}{2} \sum_{k\in \mathcal{S}_i}\sum_{j=1,j\neq i}^r \left( n_j \Delta \delta_k^2 -2 \Delta \delta_k \boldsymbol{1}_{n_j}^{\mathrm{T}}\Delta \boldsymbol{\delta}_j + \Delta\boldsymbol{\delta}_j^{\mathrm{T}} \Delta\boldsymbol{\delta}_j \right) \nonumber\\
&= \frac{1}{2} \sum_{j=1,j\neq i}^r \left( n_j \Delta \boldsymbol{\delta}_i^{\mathrm{T}}\Delta \boldsymbol{\delta}_i -2 \Delta \boldsymbol{\delta}_i^{\mathrm{T}} \boldsymbol{1}_{n_i} \boldsymbol{1}_{n_j}^{\mathrm{T}}\Delta \boldsymbol{\delta}_j + n_i \Delta\boldsymbol{\delta}_j^{\mathrm{T}} \Delta\boldsymbol{\delta}_j\right) \nonumber\\
& = \frac{1}{2}[(n-2n_i)\Delta\boldsymbol{\delta}_i^{\mathrm{T}} \Delta\boldsymbol{\delta}_i + n_i \Delta \boldsymbol{\delta}^{\mathrm{T}} \Delta\boldsymbol{\delta} - 2(\Delta \boldsymbol{\delta}_i^{\mathrm{T}} \boldsymbol{1}_{n_i})(\boldsymbol{1}_n^{\mathrm{T}}\Delta \boldsymbol{\delta} - \boldsymbol{1}_{n_i}^{\mathrm{T}} \Delta\boldsymbol{\delta}_i)]\nonumber\\
&= \Delta \boldsymbol{\delta}^{\mathrm{T}}(\frac{n-2n_i}{2}{\mathcal I}_i^{\mathrm{intra}}) \Delta \boldsymbol{\delta} + \Delta \boldsymbol{\delta}^{\mathrm{T}} (\frac{n_i}{2}\boldsymbol{I}_{n\times n}) \Delta \boldsymbol{\delta} - \Delta \boldsymbol{\delta}^{\mathrm{T}}({\mathcal I}_i^{\mathrm{intra}} \boldsymbol{1}_{n} \boldsymbol{1}_{n}^{\mathrm{T}} (\boldsymbol{I}_{n\times n} - {\mathcal I}_i^{\mathrm{intra}}) )\Delta \boldsymbol{\delta} \nonumber\\
&= \Delta \boldsymbol{\delta}^{\mathrm{T}}[\frac{n-2n_i}{2}{\mathcal I}_i^{\mathrm{intra}}+
\frac{n_i}{2}\boldsymbol{I}_{n\times n} +
{\mathcal I}_i^{\mathrm{intra}} \boldsymbol{1}_{n} \boldsymbol{1}_{n}^{\mathrm{T}} (\boldsymbol{I}_{n\times n} - {\mathcal I}_i^{\mathrm{intra}}) 
] \Delta \boldsymbol{\delta} = \mathrm{RHS},
\end{align} 
\noindent with
\begin{align}
\bar{\mathcal L}_i^{\mathrm{inter}} = \frac{n-2n_i}{2}{\mathcal I}_i^{\mathrm{intra}}+
\frac{n_i}{2}\boldsymbol{I}_{n\times n} +
{\mathcal I}_i^{\mathrm{intra}} \boldsymbol{1}_{n} \boldsymbol{1}_{n}^{\mathrm{T}} (\boldsymbol{I}_{n\times n} - {\mathcal I}_i^{\mathrm{intra}}).
\end{align}

Thus, according to eq.(\ref{Qi_cross:eq}),
\begin{align}
{\boldsymbol x}^T{\boldsymbol Q}_i{\boldsymbol x} & = \sum_{k\in s_i}\sum_{j\in s_i \atop j>k}{(\Delta \delta_k-\Delta \delta_j)^2 +(\Delta \omega_k-\Delta \omega_j)^2} +\sum_{k\in s_i}{\Delta E_k^2} \nonumber\\
& = \begin{bmatrix}\Delta {\boldsymbol \delta}\\ \Delta {\boldsymbol \omega} \\ \Delta {\boldsymbol E}\end{bmatrix}^T
\underbrace{\begin{bmatrix}\bar {\mathcal L}_i^{\mathrm{intra}} + \bar {\mathcal L}_i^{\mathrm{inter}} & ~ &~\\
~& \bar {\mathcal L}_i^{\mathrm{intra}} + \bar {\mathcal L}_i^{\mathrm{inter}} & ~\\
~&~& \mathcal{I}_{i}^{\mathrm{intra}}\end{bmatrix} }_{{\boldsymbol Q}_i'}
\begin{bmatrix}\Delta {\boldsymbol \delta}\\ \Delta {\boldsymbol \omega} \\ \Delta {\boldsymbol E}\end{bmatrix}\nonumber\\
&={\boldsymbol x}^T({\mathcal P}^T{\boldsymbol Q'}_i{\mathcal P}){\boldsymbol x},
\end{align}
\noindent and
\begin{align}
{\boldsymbol Q}_i = {\mathcal P}^T{\boldsymbol Q'}_i{\mathcal P}
\end{align}


\section{Efficiency of the grand coalition}


Consider a coalitional game where the players within each coalition cooperate while different coalitions compete. Given a coalitional structure $\rho = \{\mathcal S_1, \mathcal S_2, ..., \mathcal S_l\}$ and a set of players $\mathcal N=\{1,...,r\}$, $\rho$ is defined as a {\it partition} if $\forall i\neq j$, $\mathcal S_i \cap \mathcal S_j= \phi$, and $\cup_{i=1}^{l}\mathcal S_i=\mathcal N$ \citelatex{Saad:2010aa}. In the multi-agent control problem, the control objective of each coalition $\mathcal S \subset \rho$ is denoted as $J_{\mathcal S}$, given by
\begin{equation}
\label{Jsi:eq}
J_{\mathcal S}(\boldsymbol K^{\mathcal S},\boldsymbol K^{-\mathcal S}) = \int_{t=0}^{\infty}{[\boldsymbol x^{\mathrm{T}}(t)\boldsymbol Q_{\mathcal S}\boldsymbol x(t) + \sum_{j\in\mathcal S}\boldsymbol u_j^{\mathrm{T}}(t)\boldsymbol R_j\boldsymbol u_j(t)]dt}
\end{equation}
\noindent where $\boldsymbol K^{\mathcal S}$ is the submatrix of the feedback matrix $\boldsymbol K$ that represents the strategy of the coalition $\mathcal S$ and is given by the union of the submatrices $\boldsymbol K^j$ in (\ref{K^i:eq}) associated with agents $j\in \mathcal S$. Under the sparsity constraint $s$, the Nash strategies of the coalitions in $\rho$ are expressed as 
\begin{align}
&J_{\mathcal S}({{\boldsymbol K}^{\mathcal S}}^*,{{\boldsymbol K}^{-\mathcal S}}^*) \leq J_{\mathcal S}({\boldsymbol K}^{\mathcal S},{{\boldsymbol K}^{-\mathcal S}}^*) ~,\forall {\boldsymbol K}^{\mathcal S}\nonumber\\
\mbox{s.t. } & \mathrm{card}_{\mathrm{off}}({\boldsymbol K}) \leq s 
\label{NE_coal:eq}
\end{align}

\noindent Suppose $\boldsymbol K_{\rho} = ({\boldsymbol K^{\mathcal S_1}}^*, {\boldsymbol K^{\mathcal S_2}}^*,...,{\boldsymbol K^{\mathcal S_l}}^*)$ is the feedback matrix when the strategies of the coalitions in $\rho = \{\mathcal S_1,...,\mathcal S_l\}$ are at a Nash Equilibrium.

The value of a coalition $\mathcal S$ in the partition $\rho$ is defined as the objective reduction of $\mathcal S$, with respect to the decoupled game, i.e.
\begin{equation}
v_{\rho}(\mathcal S) = \sum_{i\in \mathcal S}{J_i^{D}} - J_{\mathcal S}(\boldsymbol K_{\rho}).
\label{v_S:eq}
\end{equation}

\noindent The above coalitional game is in partition form \cite{Hafalir2007242} since the value of each coalition depends on the composition of other coalitions. It is shown in \cite{Hafalir2007242} that for coalitonal games in partition form, the grand coalition $\mathcal N=\{1,...,r\}$ forms when it is efficient, i.e., for any partition $\rho$, the value of $\mathcal N$ is not exceeded by the combined values of the coalitions in $\rho$:
\begin{equation}
v_{\mathcal N}(\mathcal N) \geq \sum_{\mathcal S \subset \rho}v_{\rho}(\mathcal S),~\forall \rho.
\label{eff:eq}
\end{equation}

\noindent Next, suppose the matrices $\boldsymbol Q_{\mathcal S}$ in (\ref{Jsi:eq}) satisfy
\begin{equation}
 \sum_{\mathcal S \subset \rho}\boldsymbol Q_{\mathcal S} = \boldsymbol Q,
 \label{Qs:eq}
\end{equation}
\noindent which is a coalition-level equivalent of (\ref{sumQ:eq}). Then, for any partition $\rho$ of $\mathcal N$, the sum of the values of the coalitions in $\rho$
\begin{equation}
\sum_{\mathcal S \subset \rho}v_{\rho}(\mathcal S) = \tilde J^{\mathrm{D}} - J(\boldsymbol K_{\rho}) \leq \tilde J^{\mathrm{D}} - J(\boldsymbol K_{\mathcal N}) = v_{\mathcal N}(\mathcal N)
\label{ineq:eq}
\end{equation}

\noindent where $\boldsymbol K_{\mathcal N}$ is the feedback matrix that satisfies the social optimization (\ref{sparse_cent:eq}). To prove (\ref{ineq:eq}), note that $\sum_{\mathcal S \subset \rho} J_{\mathcal S}(\boldsymbol K_{\rho})=J(\boldsymbol K_{\rho})$ when (\ref{Qs:eq}) holds, and thus $\boldsymbol K_{\rho}$ represents a suboptimal solution to (\ref{sparse_cent:eq}) under the constraint $s$, resulting in $J(\boldsymbol K_{\rho}) \geq J(\boldsymbol K_{\mathcal N})$. Therefore, for any partition $\rho$, the value of the grand coalition is at least as large as the sum of the values of the coalitions in $\rho$, i.e., (\ref{eff:eq}) holds, and the grand coalition is efficient, which guarantees the formation of the grand coalition in the cooperative game and justifies Step 1 of Alg. 3 (social optimization) under the assumption (\ref{Qs:eq}).

\bibliographystylelatex{apalike}
\bibliographylatex{ref}


\end{document}